\documentclass[prd,aps,twocolumn,secnumarabic,balancelastpage,amsmath,amssymb,nofootinbib,hyperref=pdftex,longbibliography]{revtex4-2}

\usepackage{soul}           
\usepackage{color}         
\usepackage{graphics}      
\usepackage[pdftex]{graphicx}      
\usepackage{epsf}          
\usepackage{bm}            
\usepackage{verbatim}			
\usepackage{listings}       
\usepackage{caption}
\usepackage{subcaption}

\usepackage{multirow}
\usepackage{booktabs}
\usepackage{float}

\usepackage[colorlinks=true]{hyperref}  

\newcommand{\lcdm}{\ensuremath{\Lambda\mathrm{CDM}}}

 \renewcommand{\vec}[1]{\mathbf{#1}}
 
 \def\be{\begin{equation}}
\def\ee{\end{equation}}
\def\ba{\begin{eqnarray}}
\def\ea{\end{eqnarray}}

\renewcommand{\Re}{\operatorname{Re}}
\renewcommand{\Im}{\operatorname{Im}}

 \newcommand{\apjs}{Astrophys. J. Supp.}
 \newcommand{\apjl}{Astrophys. J. Lett.}
 \newcommand{\aap}{Astron. Astrophys.}

\begin{document}
\title{Circumventing Cosmic Variance via Remote Quadrupole Measurements}
\author{Arsalan Adil}
\email{aadil@ucdavis.edu}
\affiliation{Center for Quantum Mathematics \& Physics and Department of Physics \& Astronomy\\ University of California, Davis, One Shields Ave, Davis CA.}

\author{Reid Koutras}
\email          {reid.koutras@richmond.edu}
\author{Emory F. Bunn}
\email          {ebunn@richmond.edu}
\affiliation{Physics Department, University of Richmond, Richmond, VA.}


\begin{abstract}
A number of important cosmological questions can be addressed only by probing perturbation modes on the largest accessible scales. One promising probe of these modes is the Kamionkowski-Loeb effect, i.e., the polarization induced in the cosmic microwave background (CMB) by Thomson scattering in galaxy clusters, which is proportional to the CMB quadrupole measured at the cluster's location and look-back time. We develop a Fisher formalism for assessing the amount of new information that can be obtained from a future remote quadrupole survey. To demonstrate the constraining power of such a survey, we apply our formalism to a model that suppresses the primordial power spectrum on large scales but is poorly constrained with existing CMB data. We find that the constraints can be improved by over $3\sigma$ for a survey that measures around 100 clusters over $20\%$ of the sky with a signal-to-noise ratio of $3$. In the most optimistic case with a low-noise survey with dense full-sky coverage and only a single degree of freedom in the theory, the constraint improves to over $7\sigma$ beyond local CMB data. Our formalism, which is based in real space rather than harmonic space, can be used to explore a wide range of survey designs, and our results paint an optimistic picture for the utility of remote quadrupole measurements to probe physics on the largest observable scales in the Universe.

\end{abstract}

\maketitle


\section{Introduction}
\label{sec:intro}
Observations of the cosmic microwave background (CMB) provide some of the most important lines of evidence in support of the current cosmological paradigm (\textit{e.g.}, \cite{cobe,wmapresults,planckresults,spt,bicep}). Over a wide range of angular scales, the CMB anisotropy appears consistent with a realization of a Gaussian random process, with a power spectrum that matches the predictions of a $\Lambda$CDM model. 
There have, however, been multiple claims of ``anomalies'' on large angular scales that seem to be in tension with the standard model (see \cite{Schwarz_2016_CMBanomalies, bennett2011seven_cmbanomalies} for extensive reviews). Some (e.g., the alignment of low-order multipoles and the dipolar modulation of power) appear to cast doubt on bedrock assumptions of statistical homogeneity and isotropy and/or Gaussianity.

These anomalies are formally fairly significant, with $p$-values  $\sim 10^{-2}$ or lower. However, the interpretation of these significances is disputed, in large part because they are calculated \textit{a posteriori}: the anomalies were first noticed in the data, and statistics were subsequently constructed to quantify their improbability. In any large dataset, purely by chance, \textit{some} unlikely things are bound to happen, and quantifying their improbability without considering the many other unlikely things that didn't happen can be misleading. (This observation is sometimes called the ``look-elsewhere" effect.)

The way to resolve this problem is to gather new data, for which the analysis can proceed \textit{a priori}. However, the large-scale CMB anisotropy has already been measured with high signal-to-noise (except for the part of the sky that must be masked to due Galactic contamination), so further CMB temperature measurements are of no avail. We can, however, seek other methods to probe fluctuations on similar length scales. If these new observations probe different modes from the CMB, then they provide independent information, which can in principle allow us to decide if the anomalies are mere flukes or are signs of new physics. This is particularly important as several early-universe theories diverge from the standard cosmology precisely in the cosmic variance regime. This is because they modify physics at high energy scales, e.g. due to effects of quanitzed gravity or UV completion, leaving imprints on the earliest of the observable Fourier modes that left the inflationary horizon \cite{silverstein2008monodromy, mcallister2010gravity_monodromy, kaloper2011ignoble, adil2023entanglement} (see also \cite{Chluba_2015_subodhpatel, akrami2020planck} and references therein for a collection of several such models). Given that observing such signatures could lead to a paradigm shift in our understanding of modern physics, it is worthwhile to pursue avenues for constraining these theories of the early universe.

In this article, we explore one such avenue that can provide relevant information on the largest scales and study its constraining power for CMB anomalies and early universe theories. This is the method due to M. Kamionkowski and A. Loeb \cite{kamionkowski1997getting} whereby CMB photons that last scatter from electrons in rich clusters of galaxies along our line of sight get polarized -- the so-called ``polarized'' Sunyaev-Zel'dovich (SZ) effect \cite{sazonov1999microwave, sunyaev1980velocity, syunyaev}. This polarization signal is proportional to the quadrupole anisotropy in the photon distribution at the cluster's location and look-back time. Thus, while in our local measurements of the CMB only the direction of photon propagation varies, such ``remote quadrupole'' measurements allow us to probe the 3-dimensional volume contained within the last-scattering surface at $z\approx1100$. In particular, each such measurement probes a last-scattering surface centered at the cluster location and tangent to our own. These measurements thus probe scales that are smaller than, but comparable to, the lowest multipoles of our own CMB. A remote-quadrupole survey can therefore provide information that ``beats'' cosmic variance on these scales \cite{Bunn2006probing, hall2014detecting_challinor, portsmouth2004analysis, seto2005probing}.

CMB S4 will not measure the remote quadrupole signal in individual clusters, although it will detect it statisically (\textit{e.g.}, by stacking many clusters for which the signals are correlated) \cite{louis2017measuring_bunn}. A dedicated survey with similar hardware that focused specifically on clusters could provide far more information.

In recent years, several authors have studied the constraining power of remote quadrupole surveys for various cosmological phenomena includng reionization \cite{meyers2018beyond_reionization} , cluster optical depth profiles\cite{louis2017measuring_bunn}, cosmic birefringence \cite{lee2022probing_birefringence, namikawa2023cosmic_birefringence}, modified gravity \cite{pan2019forecasted_modified_gravity}, and CMB anomalies \cite{deutsch2018polarized_hemispheric_anisotropy, deutsch2018reconstruction, cayuso2020towards}. 

In this paper, we focus specifically on the lack of angular correlations on large scales. In particular, we consider a physical model that has been proposed to explain this lack of power, in which long-wavelength Fourier modes are suppressed \cite{contaldi2003suppressing}. This model has a single free parameter characterizing the amount of suppression  which is poorly constrained by current CMB data (being detected at the level of $\approx 2\sigma$). We calculate the Fisher information for this parameter for a variety of hypothetical remote quadrupole surveys. If there is a large increase in this quantity, then such a survey would give a strong answer to the question of whether the observed lack of large-scale CMB power is a mere fluke.  Such a model was also considered in Ref.~\cite{cayuso2020towards}. Our analysis differs from theirs and other related works in that we have performed all calculations in real space. As a result, we can vary survey parameters (\textit{e.g.}, redshift distribution, number of clusters, sky coverage) to determine the optimal design for a future survey. Compared to the harmonic space analyses in other works, which implicitly assume full and dense sky coverage, our results paint a rather promising outlook for the utility of performing remote quadrupole measurements on a relatively small number of clusters. 

We assess the degree to which a hypothetical future remote-quadrupole survey can shed light on these anomalies by computing the Fisher information for such surveys.
The Fisher information $F$ provides a natural metric for assessing how helpful a hypothetical set of observations will be in answering a particular question. In the case we consider in this paper, we will ask whether a future remote-quadrupole survey will significantly enhance our ability to detect a suppression in the amplitude of large-scale fluctuations. We can compute $F$ based on the existing CMB data alone, and again with the addition of remote quadrupole data. The increase in $F$ quantifies the knowledge gained as a result of the survey. In particular, $F^{-1/2}$ provides an estimate of the uncertainty on the measured parameter in such a survey. If the uncertainty found when the new survey is included is significantly less than the best-fit value of the parameter under consideration, then the new survey should give a clear answer.

In future work, we will generalize the formalism developed here to treat models in which statistical isotropy is broken. This will allow us to examine other anomalies such as the dipolar power modulation. 

The rest of this article is structured as follows. Section \ref{sec:rqs} presents the formalism for computing the predicted properties of the remote quadrupole signal, particularly the covariances of these signals with each other and with the local CMB, and details the method for deriving Fisher information from this data. It also presents the model that we use to examine the significance of the lack of large-scale power in the CMB and comments on the detectability of the signal being sought. Section \ref{sec:results} presents results, in which we quantify the additional information that can be gained from remote quadrupole surveys with a variety of designs. Section \ref{sec:conclusions} summarizes our conclusions. 

\section{Remote Quadrupole Signal}\label{sec:rqs}
There are several sources of polarization in SZ clusters~\cite{sazonov1999microwave}. Here, we focus on the polarization generated by the scattering of the primary CMB quadrupole off electrons in the intracluster medium. We perform these calculations in position space (rather than harmonic space) because we envision a survey with a relatively small number of clusters, perhaps covering only part of the sky, for which the harmonic-space coefficients may not be well estimated.

\subsection{Formalism}
\label{sec:formalism}
We imagine a survey of remote quadrupole measurements in clusters at locations $\mathbf{r}_1,\mathbf{r}_2,\ldots,\mathbf{r}_N$, defined in a comoving coordinate system with us at the origin. The observed linear polarization signal is proportional to the quadrupole that would be observed at the cluster location $\mathbf{r}$ and look-back time $\mathbf{\eta}$. To be specific, the temperature anisotropy of the CMB probed by an observer may be expanded, as usual, in the spherical harmonics,
\begin{equation}
    \frac{\Delta T(\mathbf{r}, \eta, \hat{\mathbf{n}})}{T} = \sum_{lm} \overline{a}_{lm}(\mathbf{r}, \eta) Y_{lm}(\hat{\mathbf{n}}) \ .
\end{equation}
Scattering of a CMB photon in this cluster induces a polarization $Q+iU$ that is proportional to the quadrupole coefficient $\overline{a}_{2-2}(\mathbf{r})$ at that location and time (henceforth we drop the explicit dependence on $\eta$). Here the bar indicates that the spherical harmonic expansion is done in a coordinate system in which the cluster is located on the $z$ axis. 

In the theories we consider, these coefficients are Gaussian random variables with zero mean. A complete description of the joint probability distribution therefore requires the covariances of the real and imaginary parts of these signals, which can be encoded in the covariance and relation matrix elements
\begin{equation}
\begin{aligned}
    \label{eq:covrelmat}
    &\Gamma^R_{jk}&\equiv&\langle \overline{a}_{2-2}(\mathbf{r}_j )\overline{a}_{2-2}^*(\mathbf{r}_k)\rangle ,\\
     &C^R_{jk}&\equiv&\langle \overline{a}_{2-2}(\mathbf{r}_j )\overline{a}_{2-2}(\mathbf{r}_k)\rangle=
     \langle\overline{a}_{2-2}(\mathbf{r}_j)\overline{a}^*_{22}(\mathbf{r}_k)\rangle    , 
\end{aligned}
\end{equation}
where the superscript `R' is to explicitly specify that these matrices encode the remote-remote correlations. 


It is easiest to begin by considering the coefficients $a_{2m}(\mathbf{r})$ in a fixed coordinate system first, and then apply the appropriate rotation operators to determine the statistics of the coefficients $\overline{a}_{2m}(\mathbf{r})$ in the  rotated coordinate system that defines the polarization basis.

The coefficient ${a}_{2m}$ can be expressed as
\begin{equation}
    a_{2m}(\vec r) = \int d^3k \,\Delta_2(k,r)\delta_\Phi(\vec k) e^{i\vec k\cdot\vec r} Y_{2m}^*(\hat{\mathbf{k}}),
\end{equation}
where $\delta_\Phi$ is the Fourier-space perturbation to the gravitational potential.
On the scales of interest, the transfer function $\Delta_2$ includes only Sachs-Wolfe and integrated Sachs-Wolfe contributions and can be written
\begin{widetext}
\begin{equation}
\label{eq:transfer-function}
    \Delta_2(k,r) = -{4\pi\over 3}
    \left(j_2[k(\eta-\eta_{\mathrm{rec}})]+
    6\int_{\eta_{\mathrm{redc}}}^\eta d\eta'\,j_2[k(\eta-\eta')]{\partial\over\partial\eta'}\left(D(\eta')\over a(\eta')\right)\right).
\end{equation}
In this expression $j_2$ is a spherical Bessel function; $a$ is the scale factor normalized to one at the present time; $\eta_{\mathrm{rec}},\eta,\eta_0$ are conformal times at recombination, the cluster look-back time, and the present respectively; and $D$ is the matter perturbation growth factor~\cite{padmanabhan2003cosmological}. We need to know covariances of the form:
\begin{equation}
\gamma_{m_1m_2}(\mathbf{r}_1,\mathbf{r}_2)\equiv \langle a_{2m_1}(\mathbf{r}_1)
    a_{2m_2}^*(\mathbf{r}_2)\rangle \ .
\end{equation}
The perturbations $\delta_\Phi$ are drawn from a homogeneous and isotropic Gaussian random process with
\begin{equation}
    \langle\delta_\Phi(\mathbf{k})
    \delta_\Phi^*(\mathbf{k}')\rangle
    =P_\Phi(k)\delta^3(\mathbf{k}-\mathbf{k}')
\end{equation}
for some power spectrum $P_\Phi$. Thus, we can write
\begin{equation}
    \gamma_{m_1m_2}(\mathbf{r}_1,\mathbf{r}_2)
    =\int d^3k\,\Delta_2(k,r_1)
\Delta_2(k,r_2)P_\Phi(k)e^{i\mathbf{k}\cdot(\mathbf{r}_1-\mathbf{r}_2)} 
Y_{2m_1}^*(\hat{\mathbf{k}})
Y_{2m_2}(\hat{\mathbf{k}}) \ .
\end{equation}
We can write this expression in a form that is more efficient to evaluate by expanding the exponential in spherical harmonics. The angular part of the triple integral is then over the product of three spherical harmonics and can be expressed in terms of Clebsch-Gordan coefficients \cite{zare}. The result is
\begin{equation}
    \gamma_{m_1m_2}(\mathbf{r}_1,\mathbf{r}_2) = 
    4\pi \sum_{L=0,2,4} i^L I_L Y^*_{LM} (\Delta \hat{r}) \langle L \, M | 2 \, m_1 \, 2 \, m_2 \rangle,
    \label{eq:covarsum}
\end{equation}
where
\begin{equation}
    I_L = \int_{0}^{\infty} dk P_\Phi(k) k^2 \Delta_2(k,r_1) \Delta_2 (k,r_2) j_L(k\,\Delta r),
    \label{eq:Iint}
\end{equation}
$M=m_1-m_2$,  $\Delta{\mathbf{r}}=\mathbf{r}_1-\mathbf{r}_2$, and $\langle LM|2m_12m_2\rangle$ denotes a Clebsch-Gordan coefficient.
\end{widetext}

We can compute covariance and relation elements for each pair of clusters by first computing the
$5\times 5$ matrix with elements $\gamma_{m_1m_2}(\mathbf{r}_1,\mathbf{r}_2)$, and then applying Wigner $D$ matrices \cite{zare} 
on both left and right to convert each set of $a_{2m}$ coefficients to the rotated coordinate system for that cluster. The $a_{lm}$ coefficients transform under the action of the Wigner matrices as
\begin{equation}
\label{eq:wigner}
    \bar{a}_{lm} = \sum_{m'} D^{l}_{mm'}({\mathcal R})a_{lm'} \ ,
\end{equation}
where $\mathcal{R}$ is the rotation that converts one coordinate system to the other.

The covariance and relation matrix elements are the $(m_1,m_2)=(-2,-2)$ and $(-2,2)$ elements of the resulting matrix.\footnote{It may seem inefficient to compute an entire $5\times 5$ matrix when we only need two elements, but the computationally expensive step is the numerical integration (\ref{eq:Iint}), which is independent of $m_1,m_2$ and so only needs to be done once (per $L$) for each matrix element.}

In addition to the covariances of remote quadrupole signals, we also need to know the cross-covariances with the locally measured CMB anisotropy. These are computed similarly: to compute $\langle a_{2m_1}(\mathbf{r}_1)a_{lm_2}^*(0)\rangle$ we set $r_2=0$ and replace the transfer function $\Delta_2$ with the appropriate $\Delta_l$. As in Eq.~(\ref{eq:covarsum}), the result can be expressed as a sum of at most three terms, with $L=l-2,l,l+2$. This is represented by: 
\begin{equation}
\begin{aligned}
\langle {a}_{2m_1}(\vec{r_1})a^*_{lm}(0) \rangle = 4\pi \sum_{L=l-2,l,l+2}^{\infty} 
\langle L \, M | 2 \, m_1 \, l \, m \rangle i^L \\
\times \int_{0}^{\infty} dk P_\Phi(k) k^2  \Delta_2(k,\vec{r_1}) \Delta_l^* (k,0) j_L(kr) Y^*_{LM}(r) \ . 
\label{eq:cross}
\end{aligned}
\end{equation}
The Clebsch-Gordan coefficients show that we only get a nonzero answer if $M=m_1-m$, so we no longer have to sum over m. The above equation is derived in a similar way to    Eq. (\ref{eq:covarsum}).

As in the remote case, we wish to convert $a_{2m}(\mathbf{r}_1)$ from the fixed coordinate system to the appropriate rotated coordinate system, which we can do by applying the appropriate Wigner matrix (Eq.~\ref{eq:wigner}). We then pick out the coefficients $m=\pm 2$, which correspond to the observable signal. (The local coefficients $a_{l m}$ remain in the fixed coordinate system.)

\subsection{Methodology}
\label{sec:methods}
In the theories we will consider, the observed data (remote quadrupole measurements and local CMB anisotropy) are drawn from a multivariate normal distribution with zero mean. We will represent the data as a vector $\vec d$. For the moment, we take $\vec d$ to be real rather than complex; we will explicitly show the changes required for complex observables below. 

Under these assumptions, a complete description of the probability distribution requires only the covariance matrix $\mathbf{\Psi}\equiv\langle \vec d\vec d^T\rangle$.  

We first consider a family of theories parameterized by a parameter $p$. The \textit{Fisher information } (\textit{e.g.}, \cite{berger}) is then defined as $F=-\partial^2(\ln L)/\partial p^2$, where $L$ is the likelihood. For the Gaussian theories under consideration, this is
\begin{equation}
    F = {1\over 2}\mbox{Tr}(
(\mathbf{\Psi} ^{-1}\mathbf{\Psi}')^2,
    \label{eq:fi}
\end{equation}
where $\mathbf{\Psi}'=\partial\mathbf{\Psi}/\partial p$.
The Cram\'er-Rao inequality says that any unbiased estimator of $p$ must have standard deviation
\begin{equation}
    \sigma_{p} \ge F^{-1/2}.
    \label{eq:cramerrao}
\end{equation}

Although there is no general guarantee of this, in practice this inequality is often close to equality for the optimal estimator. Following common practice, we interpret $F$ in this way in this paper. Naturally, if an experimental design appears promising by this metric, one should perform a more detailed calculation (e.g., simulation) to see how close the actual errors are to the Fisher bound.

If the family of theories under consideration contain multiple parameters $p_1,p_2,\ldots$, the Fisher information is characterized by a matrix with elements $F_{ij}=-\partial^2(\ln L)/\partial p_i\partial p_j$. The Cram\'er-Rao bound on the uncertainty in a particular parameter, assuming the others are unknown, is then
\begin{equation}
    \sigma_{p_j}\ge (\mathbf{F}^{-1})_{jj}^{-1/2} \ .
\end{equation}
 If the other parameters are known, then the bound is
\begin{equation}
    \sigma_{p_j}\ge (F_{jj})^{-1/2} \ .
\end{equation}
The former equation corresponds to marginalizing over the other parameters $p_i$ with $i\ne j$, while the latter comes from holding them fixed at given values.

If one parameter $p_1$ is of primary interest, and the others are considered to be nuisance parameters, then it is convenient to define the Fisher information $F$ for that one parameter to be that which makes the one-parameter Cram\'er-Rao inequality (\ref{eq:cramerrao}) true. To be specific, we define the scalar Fisher information in this case to be 
\begin{equation}
    F=1/(\mathbf{F}^{-1})_{11}
    \label{eq:fisherfixednuisance}
\end{equation}
when we are marginalizing over the other parameters, and $F=F_{11}$ if the other parameters are treated as known.

Our primary interest will be in the question of whether we can be confident that the parameter of interest $p_1$ is nonzero (\textit{i.e.}, in whether the null hypothesis that $p_1=0$ can be rejected). In such cases, it is convenient to refer to the normalized Fisher information, 
\begin{equation}
    F_n = \hat p_1^2 F,
\end{equation}
where $\hat p_1$ is the best-fit value of the parameter. The null hypothesis is rejected when $F_n\gg 1$. For notational convenience, we refer to this normalized Fisher information as just the Fisher information and suppress the subscript `$n$' in the rest of this paper.

Because we are interested in the additional information obtainable by a future survey, and because the future data are correlated with existing measurements of the local CMB, we are particularly interested in the increase in Fisher information when the new survey data are added to the previously-known CMB data.

Our data will consist of observed remote quadrupole signals $\overline{a}_{2-2}(\mathbf{r})$ for a variety of cluster locations. 
We wish to consider a variety of possible surveys, covering different redshift ranges, regions of sky, etc. To efficiently explore this parameter space, we divide the Universe up into voxels covering the range $0<z\le 2$ and compute the full covariance matrix for all  voxels. 

We choose voxels to be small enough that neighboring voxels are 
highly correlated and provide redundant information. Specifically, we demand that neighboring voxels are at least 96\% correlated with each other. This criterion leads us to select 14 redshift shells $z\in \{$0.07, 0.14, 0.22, 0.31, 0.42, 0.52, 0.65, 0.79, 0.95, 1.1, 1.3, 1.5, 1.7, 2.0$\}$. We then use HEALPix \cite{healpix} to pixelize each $z-$shell  and find that $N_{\mathrm{side}} = 8$ matches our correlation criterion in all redshift bins. Thus, each $z-$shell contains $12\times N_{\mathrm{side}}^2 = 768$ pixels for a total of $N_c = 14\times 768 = 10752$ voxels. 

For these voxel locations, we compute the $N_c \times N_c $ remote covariance and relation matrices, $\mathbf{\Gamma}^R$ and $\mathbf{C}^R$ respectively [Eq.~(\ref{eq:covrelmat})]. We will also need to include the coefficients $a_{lm}$ of the local CMB in $\mathbf{d}$, because the remote quadrupole measurements are correlated with them. We must therefore also compute the
the $N_c\times N_b$ matrices of cross-correlations
$\mathbf{\Gamma}^X$ and $\mathbf{C}^X$ [Eq.~(\ref{eq:cross})] using the fiducial cosmology of Section~\ref{sec:fiducial-model}. Here, $N_b = \sum_{l=2}^{l_{\rm max}} (2l+1) = l_{\rm max}^2 +2l_{\rm max} -3$ since each  $l$  has $2l + 1$ independent azimuthal $m$ modes. 

The reason one must consider these remote-local cross-correlations is that the Fourier modes that project onto the quadrupole probed by remote observers in fact project onto the higher multipoles at $l > 2$ for the CMB observed by us today (recall that the last-scattering surface probed by the remote observers is tangent to, and smaller than, our last-scattering surface). In principle, the cross-correlation with the local temperature anisotropy may extend to arbitrarily large $l$ but, for the redshift range relevant for the remote quadrupole signal, one need only consider the first few multipoles. We find that for the highest-redshift clusters in our sample (at $z=2$), the cross-correlation with the local CMB is negligible beyond $l=6$. 

One must also account for the information contained in the local CMB by calculating the angular power spectrum $C_l$ up to $l_{\rm max}$. We use these to create the $N_b\times N_b$ covariance and relation matrices $\mathbf{\Gamma}^L$ and $\mathbf{C}^L$ for the local CMB. The former is diagonal, with $2l+1$ copies of each $C_l$. The latter is similarly sparse, containing off-diagonal elements corresponding to $\langle a_{lm}a_{l-m}\rangle=(-1)^mC_l$.

Finally, to account for the effect of noise on these measurements, we add a noise covariance matrix $\mathbf{N}^R$ to $\mathbf{\Gamma}^R$. We assume uncorrelated isotropic errors so that $\mathbf{N}^R$ is parameterized by a single real number,
\begin{equation}
    \label{eq:noisecov}
    N^R_{ij} = \langle n(\mathbf{r}_i),n(\mathbf{r}_j)\rangle  = \epsilon^2 \Gamma_{ii}^R\delta_{ij} \ , 
\end{equation}
where $\epsilon$ is the fractional error in each remote quadrupole measurement.

If we take our data vector $\mathbf{d}$ to be the concatenation of $N_b$ local CMB coefficients and $N_c$ remote quadrupole measurements, we can then combine all these pieces into the full hermitian covariance matrix,

\begin{equation}
\mathbf{\Gamma}=
\left[\begin{matrix} \mathbf{\Gamma}^L & \mathbf{\Gamma}^X\\
(\mathbf{\Gamma}^X)^\dag & \mathbf{\Gamma}^R\end{matrix}\right]
\end{equation}
The construction for the full complex \textit{symmetric} relation matrix $\mathbf{C}$ proceeds similarly. 

Finally, we can combine all of these ingredient into the real-valued covariance matrix, 
\begin{equation}
{\bf \Psi} =
  \begin{bmatrix}
    \Psi_{xx} & \Psi_{xy}  \\
    \Psi_{yx} & \Psi_{yy}
  \end{bmatrix}
\end{equation}
where 
\begin{eqnarray*}
\Psi_{xx} = \frac{1}{2}(\Re({\bf \Gamma}) + \Re({\bf C}) ),\\
\Psi_{xy} = \frac{1}{2}(\Im({\bf C}) - \Im({\bf \Gamma}) ),\\
\Psi_{yx} = \frac{1}{2}(\Im({\bf \Gamma}) + \Im({\bf C}) ),\\
\Psi_{yy} = -\frac{1}{2}(\Re({\bf C}) - \Re({\bf \Gamma}) ),
\end{eqnarray*}
(e.g., \cite{liubunn}), and use Eq.~(\ref{eq:fi}) to find the Fisher information on the relevant parameter.

\subsection{Fiducial Model}
\label{sec:fiducial-model}
\begin{figure}
    \centering
    \begin{subfigure}{0.45\textwidth}
        \includegraphics[width=\linewidth]{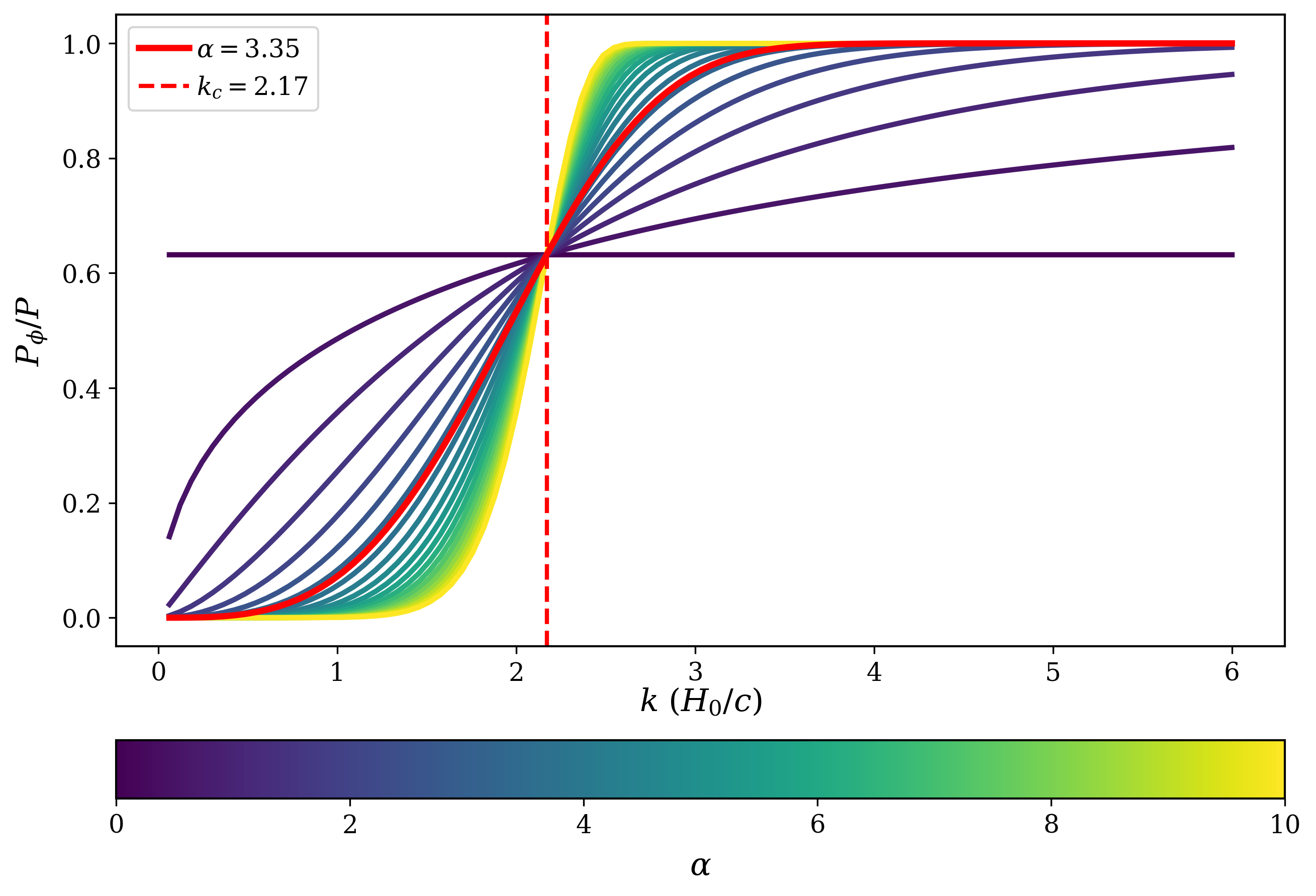}
    \end{subfigure}
    \vspace{-0.3\baselineskip}
    \begin{subfigure}{0.45\textwidth}
        \includegraphics[width=\linewidth]{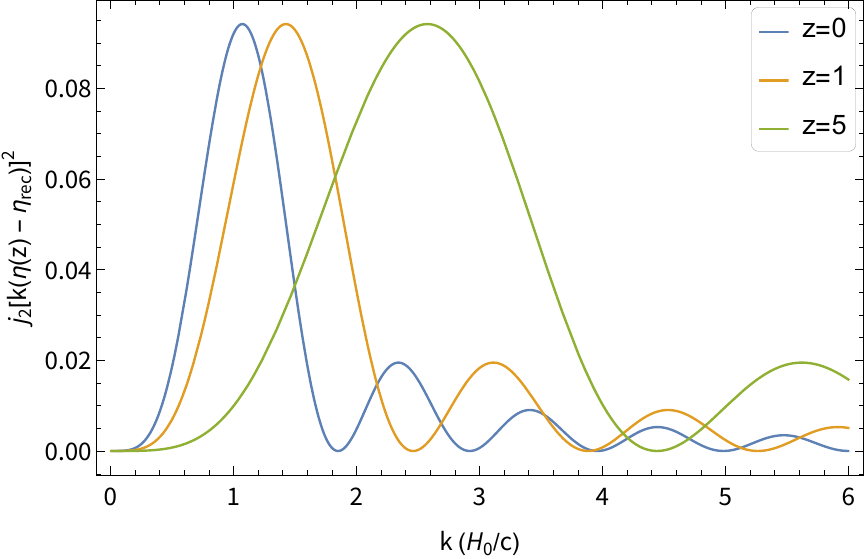}
   \end{subfigure}
    \vspace{-0.3\baselineskip}
    \begin{subfigure}{0.45\textwidth}
        \includegraphics[width=\linewidth]{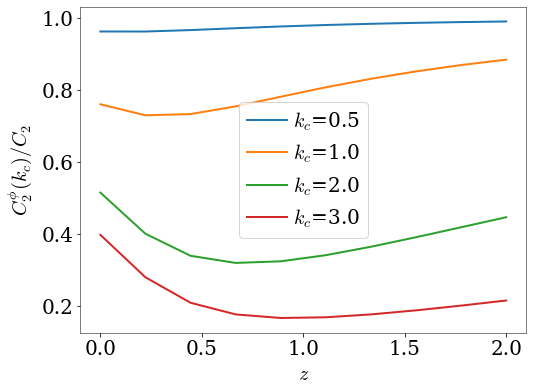}
    \end{subfigure}
    \caption{\textit{Top:} The primordial power spectrum of the fiducial model [Eq.~(\ref{eq:ps})] relative to the standard nearly scale-invariant spectrum. The red dashed line shows the best-fit value of $k_c$, while the red solid curve is the result of fixing $\alpha$ to its best-fit value. \textit{Middle:} The Sachs-Wolfe term of the transfer function [Eq.~(\ref{eq:transfer-function})] for inferring the sensitivity of the remote quadrupoles to different Fourier modes at various redshifts. \textit{Bottom:} The quadrupole of the fiducial cosmology, relative to \lcdm, as a function of redshift for various choices of the cutoff parameter.}
    \label{fig:toy-model}
    \vspace{-1.5\baselineskip}
\end{figure}

We wish to quantify the efficacy of a remote quadrupole survey to assess the significance of the large-scale power deficit. We choose a fiducial model for the primordial power spectrum that was first introduced by Contaldi et al.\cite{contaldi2003suppressing} to explain this defcit,
\begin{equation}
\begin{aligned}
&P_\phi(k) =  (1-\exp[-k/k_c]^\alpha) P(k) \\
&= A_s(1-\exp[-k/k_c]^\alpha)(k/k_0)^{n_s-1}, 
\label{eq:ps}    
\end{aligned}
\end{equation}
which suppresses power for modes below some cutoff scale $k_c$, compared to the standard nearly scale-invariant spectrum, as depicted in the top panel of Fig.~\ref{fig:toy-model}. Here $A_s$, $n_s$, and $k_0$ are the usual amplitude, tilt, and pivot scale parameters (respectively).  In their best-fit model to the \texttt{WMAP} data, Contaldi et al. find $k_c=2.17^{+0.57}_{-0.70}$ (in units of $H_0/c$) with  $\alpha=3.3$. The constraints from \texttt{Planck15} are consistent with these \cite{ade2016planck_inflation, Vitenti_2019_patrick}. We are primarily interested in constraints on the parameter controlling the physical cutoff scale, $k_c$. The other degree of freedom in the model, $\alpha$, controls the steepness of the drop (see the top panel of Fig.~\ref{fig:toy-model}) and we treat it as a nuisance parameter as described in Section \ref{sec:methods}. In Section~\ref{sec:results}, we show how our analysis depends on this model parameterization by both marginalizing as well as fixing $\alpha$ in calculating the constraints on $k_c$ from the simulated remote quadrupole measurements.

To understand this $2-3\sigma$ preference for the best-fit value of $k_c$ \footnote{Quantifying the exact preference using a Gaussian measure is difficult due to the non-Gaussian nature of the posterior, as well as dependence on the choice of priors \cite{contaldi2003suppressing}.}, we show the effect of the power suppression (with $\alpha$ fixed to the best-fit value) on the variance in the quadrupole at various redshifts in the bottom panel of Fig.~\ref{fig:toy-model}. In particular, for the best-fit value of $k_c=2.17$, the local quadrupole is suppressed by roughly a factor of $2$, while the suppression becomes negligible at all red-shifts for $k_c \lesssim 0.5$. This amount of suppression brings more concordance with the observed CMB quadrupole of $C_2^{\rm obs}=236^{+558}_{-139} {\rm \mu K^2}$ (cosmic variance errors) from the best-fit \lcdm\ predicted value of $C_2\approx1065{\rm \mu K^2}$. 

Inevitably, the cause for the suppression of the local quadrupole also leads to a suppression of the \textit{remote} quadrupoles as shown in the bottom panel of Fig.~\ref{fig:toy-model}. This can be understood through Eq.~(\ref{eq:transfer-function}) where, ignoring the ISW contribution (which does not appreciably contribute to the transfer function when $\Omega_\Lambda \lesssim \Omega_m$ at $z\gtrsim0.3$), 3D Fourier modes of wavelength $k$ are projected onto the 2D harmonic mode $l=2$ via $j_2[k(\eta-\eta^*])$ for observers who probe a last-scattering surface at $\eta(z)$. Therefore, the contribution to the variance in the remote quadrupole from the suppressed modes (denoted $C_2^\phi$) varies with $z$. Visually, one can ``eyeball'' the difference in the contribution to $C_2(z)$ in the standard model versus that from the power spectrum $P_\phi(k)$ of Eq.~(\ref{eq:ps}) for wavelengths below the cutoff scale by comparing the spherical Bessel function with $P_\phi(k)$ (i.e. compare the middle and top panels in Fig.~\ref{fig:toy-model}). Clearly, the broadening of the Bessel function with increasing $z$ shows that for high redshifts, the contribution to $l=2$ comes increasingly from Fourier modes where the test model is asymptotic to the standard model ($k \gtrsim k_c$). Moreover, it is precisely at $z\approx 1$ where the window function is primed to maximize the overlap with the regime where $P_\phi(k)$ digresses most from $P(k)$ (i.e. at $k\lesssim k_c$). This explains the resulting curve for $k_c = 2.0$ in the bottom panel of Fig.~\ref{fig:toy-model} where the maximal suppression in $C_2(z)$ occurs at $z\approx 1$ (the same reasoning can be extended to curves corresponding to other values of $k_c$ as well). 

None of this should come as a surprise: the model is specifically designed to improve goodness-of-fit statistics by suppressing the quadrupole and leaving other observable (particularly $l> 6$ modes) nearly unchanged. That is why this model is a promising one to consider as a candidate theory to test the efficacy of the proposed remote quadrupole measurements: it is favored by current data, involves — at least in some sense — a ``look-elsewhere effect'', and it leaves an imprint on observables that have yet to be measured.

\subsection{Signal Detectability}
\label{sec:signal-detec}
The polarization signal due to the scattering of the primary CMB quadrupole scales linearly with the optical depth of the galaxy cluster. Thus, for the Planck measurement of the local quadrupole, and assuming typical optical depths of $\mathcal{O}(10^{-2})$, this polarization signal is expected to be $p_{\pm}\sim\tau \sqrt{C_2} \sim 50 \rm{nK}$. Although CMB S4 is expected to catalogue $\mathcal{O}(10^{5})$ SZ clusters, it is unlikely that it will directly measure this polarized SZ signal with high statistical significance, given the noise and resolution specifications (see Table 2.3 in \cite{abazajian2019cmbs4_science_book}).    

For our analysis, we depict the Fisher information for a variety of noise levels. We imagine a survey with limited telescope time so that the noise per cluster is moderated by the square root of the number of clusters measured by such a survey. This noise parameter is given by
\begin{equation}
    \nu=\epsilon/\sqrt{n_c} \ ,
\end{equation}
where $\epsilon$ is the noise-to-signal ratio, $\rm{SNR}^{-1}$, for each measurement [see Eq.~(\ref{eq:noisecov})]. 

How does this noise parameter translate to instrument specifications? Let us do a back-of-the-envelope calculation based on some simplifying assumptions. The error in each cluster measurement, $e^c_i = \epsilon \sqrt{\Gamma_{ii}}$, can be written in terms of the pixel-level detector noise as
\[
e^c_i = \frac{N_{\rm det}}{\sqrt{t_c} } \frac{\theta^p}{\theta^c_i} = \frac{N_{\rm det}}{\sqrt{T_s} } \frac{\theta^p}{\theta^c_i} \sqrt{n_c} \ ,
\]
where $t_c$ is the integration time over the cluster, $\theta^p$ is the angular resolution, $\theta^c$ is the angular extent of the signal, and $T_s$ is the total survey time. Here we have absorbed any dependence on the number of detectors into $N_{\rm det}$ and on the survey efficiency into $T_s$. Thus, $\nu$ is related to the detector specifications by
\[
\nu \sqrt{\Gamma_{ii}} = \frac{N_{\rm det}}{\sqrt{T_s} } \frac{\theta^p}{\theta^c_i} \ .
\]

Let us now do a rough approximation for the instrument specifications given some desired SNR on our measurements. Suppose we wish to measure $n_c=100$ clusters, each with a SNR of $1.0$, so that $\nu=0.1$. For an approximation, let us assume that the signal is constant across the cluster, and is the same ($\approx 50$ nK for all clusters), with typical angular extents of $10 \ \rm{arcmin}$. Then, for a hypothetical survey with $T_s \approx 10^7s$ and $\theta^p \approx 1 \ \rm{arcmin}$, the requisite noise-equivalent temperature sensitivity is $N_{\rm det} \approx 150\mu \mathrm{K \sqrt{s}}$. 

This is close to the sensitivities of current and upcoming surveys such as SPT and CMB-S4. We emphasize that this is merely an order-of-magnitude estimate: systematic errors will play a significant role for a signal that is at the noise limit.
Particular challenges to detection involve uncertainties in modeling the optical depth profiles \cite{hall2014detecting_challinor}, and a plethora of other sources of secondary polarization \cite{sazonov1999microwave, sunyaev1980velocity}. The most notable of these confusion-sowing sources are those generated due to the single scattering of CMB photons in a cluster moving transverse to our line of sight $\propto \tau \beta_t^2$, and two sources due to the secondary scatter of unpolarized CMB radiation in a cluster with finite optical depth, $\propto \tau^2 \beta_t$ and $\tau^2 kT_e/m_e c^2$. While these effects are roughly the same order of magnitude as the polarization due to the scattering of the primary quadrupole, $\mathcal{O}(10^{-8})\rm{K}$, they can be separated, in principle, due to their spatial and spectral signatures, as well as differences in coherence lengths. A full map-making scheme is beyond the scope of this article but we refer the reader to Refs.~\cite{sazonov1999microwave, sunyaev1980velocity} for an analytic overview of the formalism, Refs.~\cite{shimon2006cosmic_raphaeli, amblard2005sunyaev_white, mirmelstein2020detection} for simulations depicting spatial variations of the polarization across a cluster, and Ref.~\cite{shimon2009power} for the frequency dependencies of the various effects. We also refer the reader to Ref. \cite{khabibullin2018polarization} for an overview of other subdominant sources of secondary polarization (see Table 3 therein for a summary).
Further hope for an early detection comes from the pSZ tomography method laid out in Ref.~\cite{deutsch2018reconstruction} which can significantly improve the signal-to-noise ratio by correlating the SZ polarization signal with the distribution of tracers of large scale structure. Moreoever, one can get a handle on errors in the optical depth profiles of SZ clusters by measurements of the usual thermal and kinematic SZ effects generated by these clusters (see e.g. \cite{schiappucci2023measurement_of_tau} for a recent implementation of this technique using SPT-3G and DES data) and further via the correlated part of the polarized SZ emission, as in the approach of Ref.~\cite{louis2017measuring_bunn}. 

\section{Results \& Discussion}
\label{sec:results}

\begin{figure}
    \includegraphics[width=0.48\textwidth]{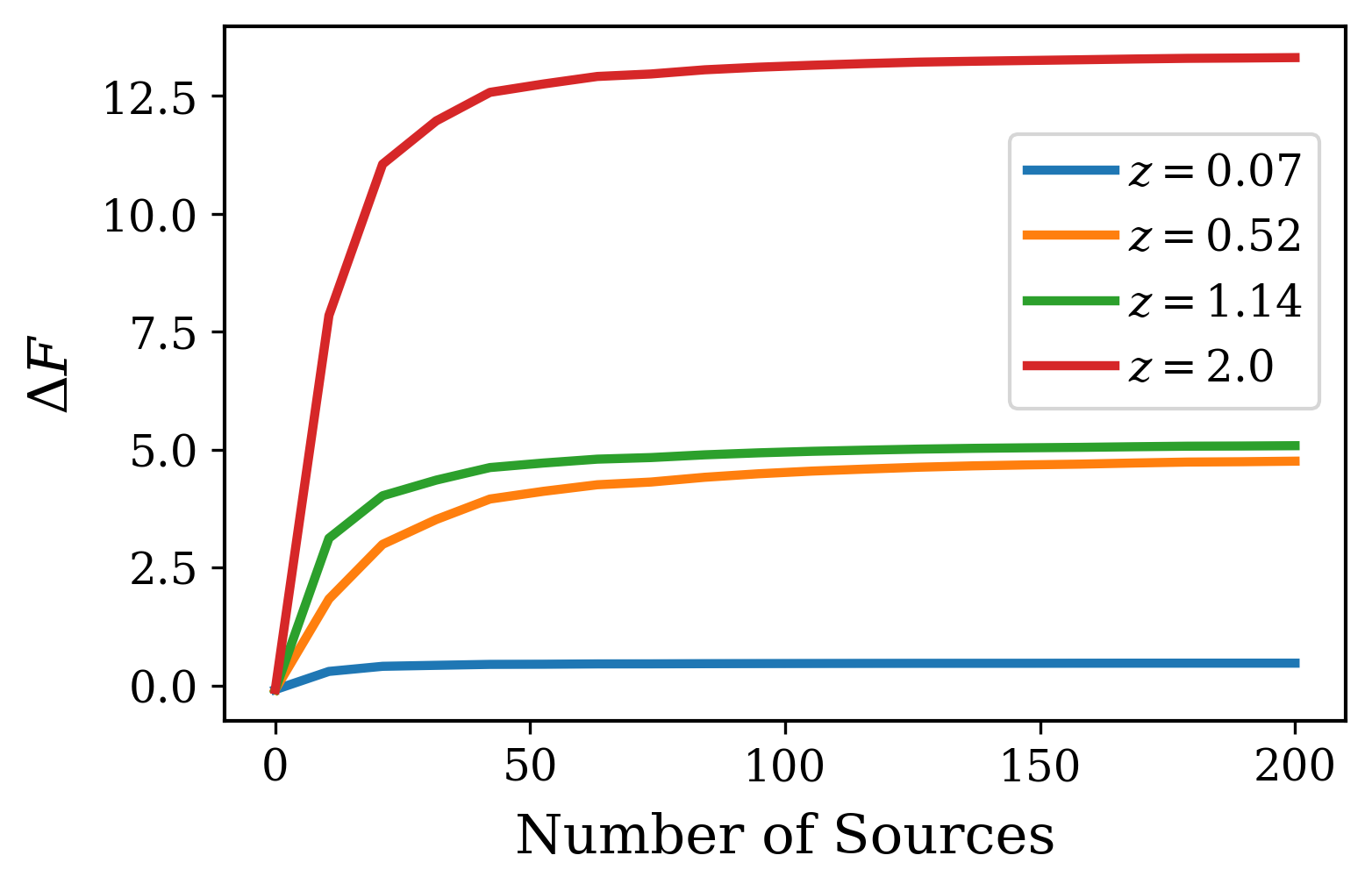}
     \caption{The increase in the normalized Fisher information of $k_c$ as a function of the number of clusters measured at a fixed redshift. We choose a representative sample of four (out of 14) redshift shells to demonstrate the variation in $F$ with $z$. The noise parameter is fixed to $\nu=0.03$ for all cases. We have truncated the plot at $N=200$ since $F$ sufficiently plateaus, for all redshits, by this point.}
    \label{fig:FvsN} 

\end{figure}

\begin{figure*}
    \centering
    \begin{subfigure}{0.35\textwidth}
        \centering
        \large{$f_{\rm sky}=0.2$}
        \includegraphics[width=\linewidth]{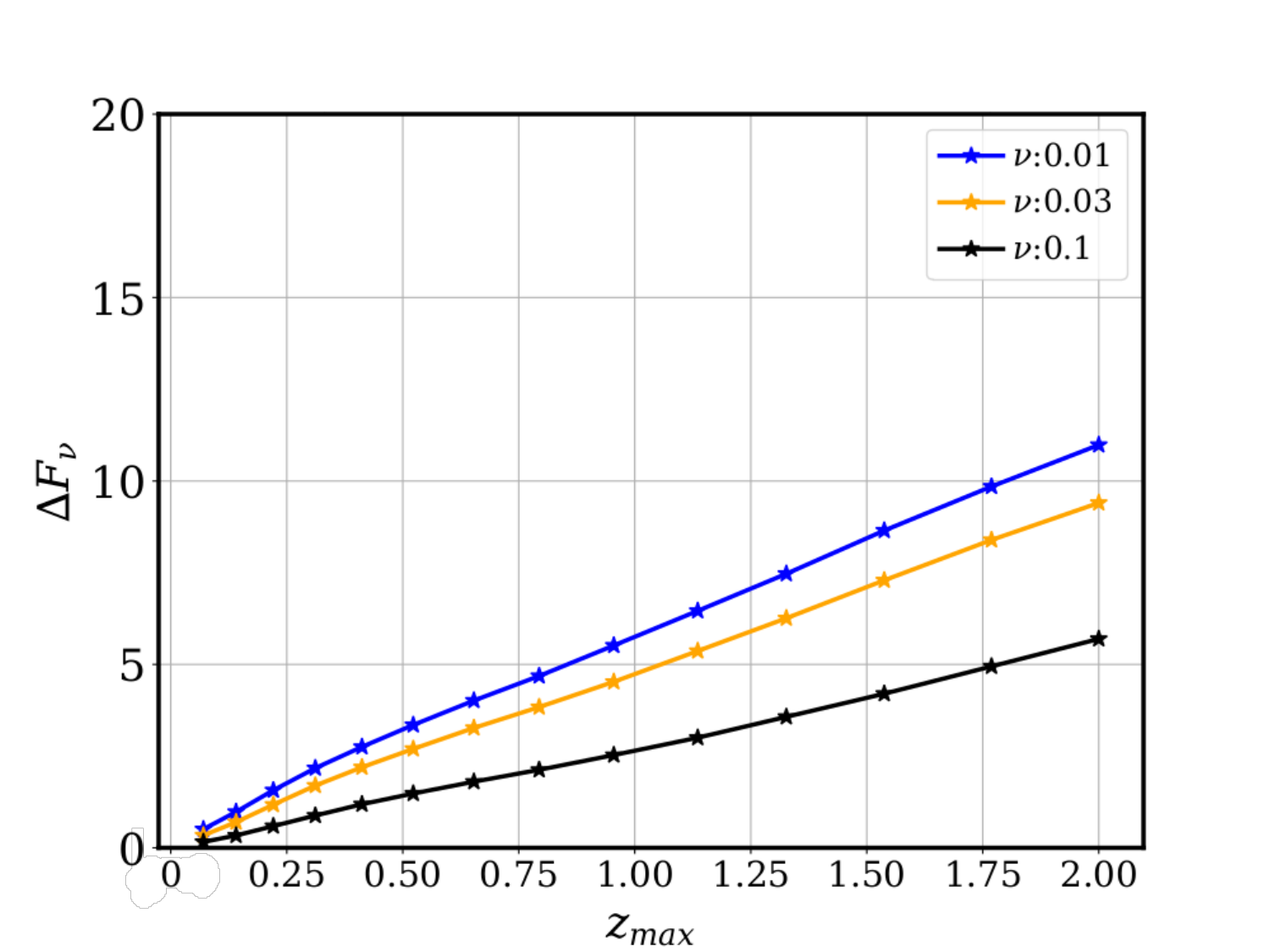}
    \end{subfigure}
    \hspace{-0.7cm}
    \begin{subfigure}{0.35\textwidth}
        \centering
        \large{$f_{\rm sky}=0.5$}
        \includegraphics[width=\linewidth]{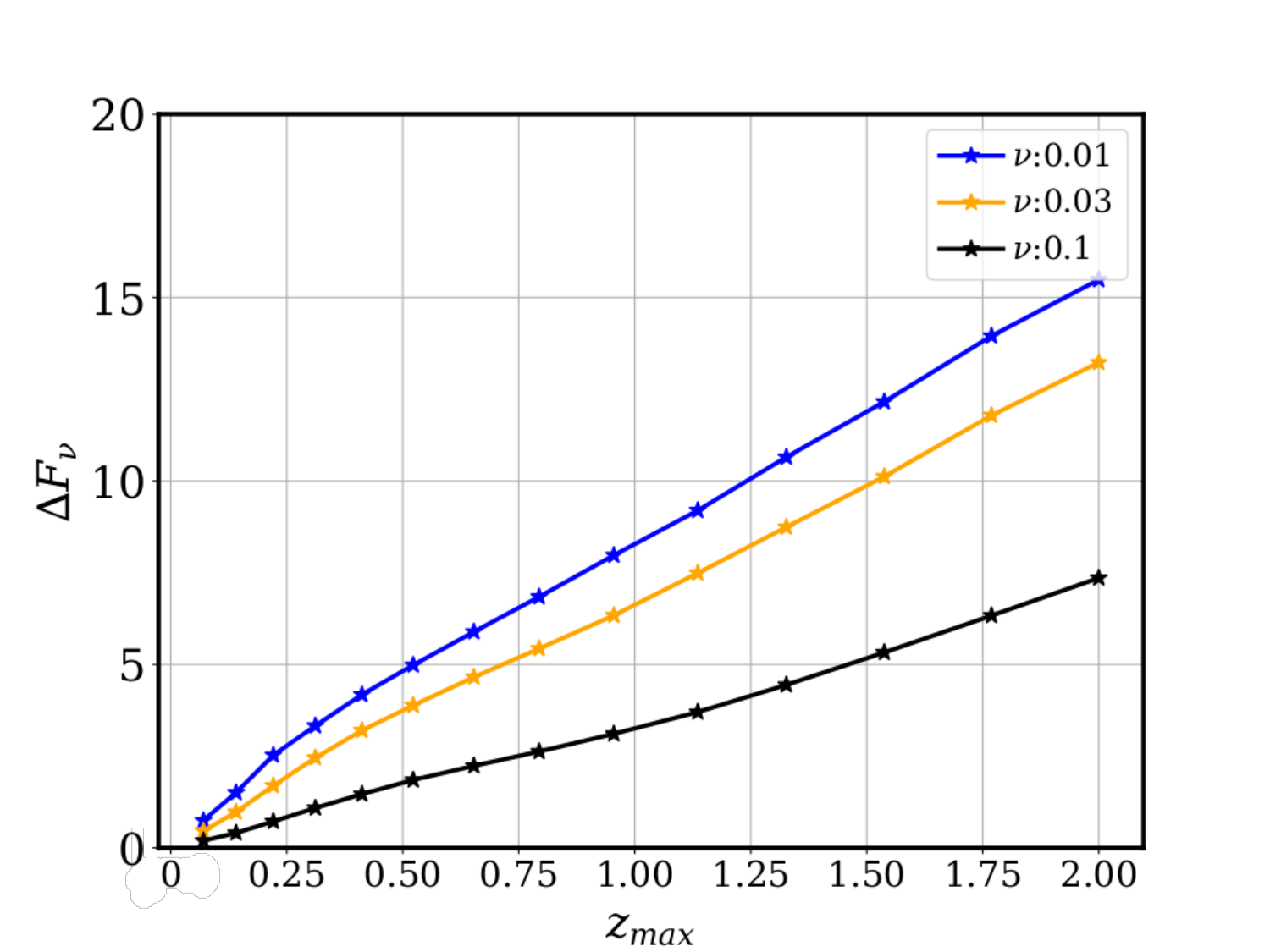}
    \end{subfigure}
    \hspace{-0.7cm}
    \begin{subfigure}{0.35\textwidth}
        \centering
        \large{$f_{\rm sky}=1.0$}
        \includegraphics[width=\linewidth]{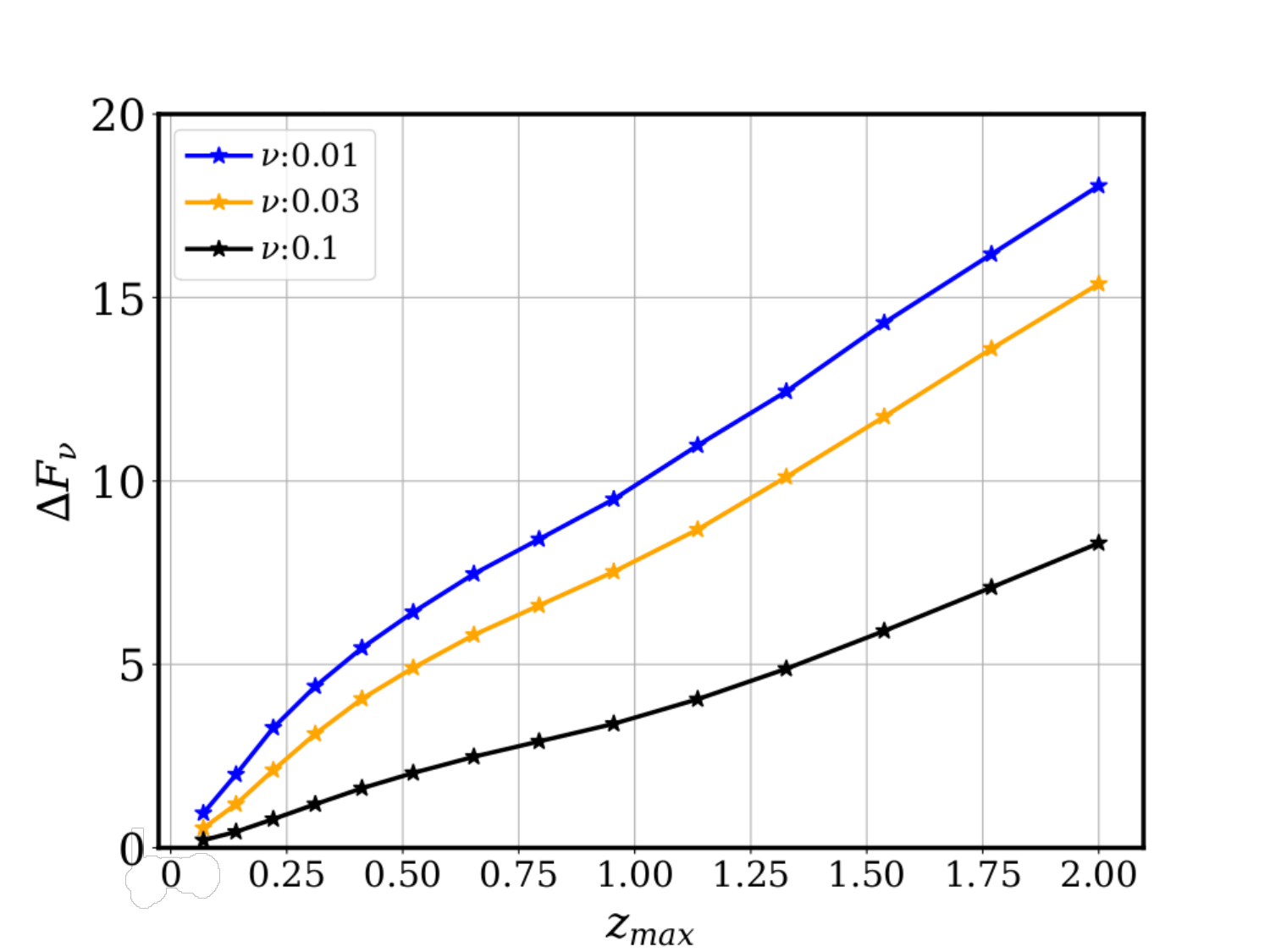}
    \end{subfigure}
    \vspace{1cm}
    \begin{subfigure}{0.35\textwidth}
        \centering
        \includegraphics[width=\linewidth]{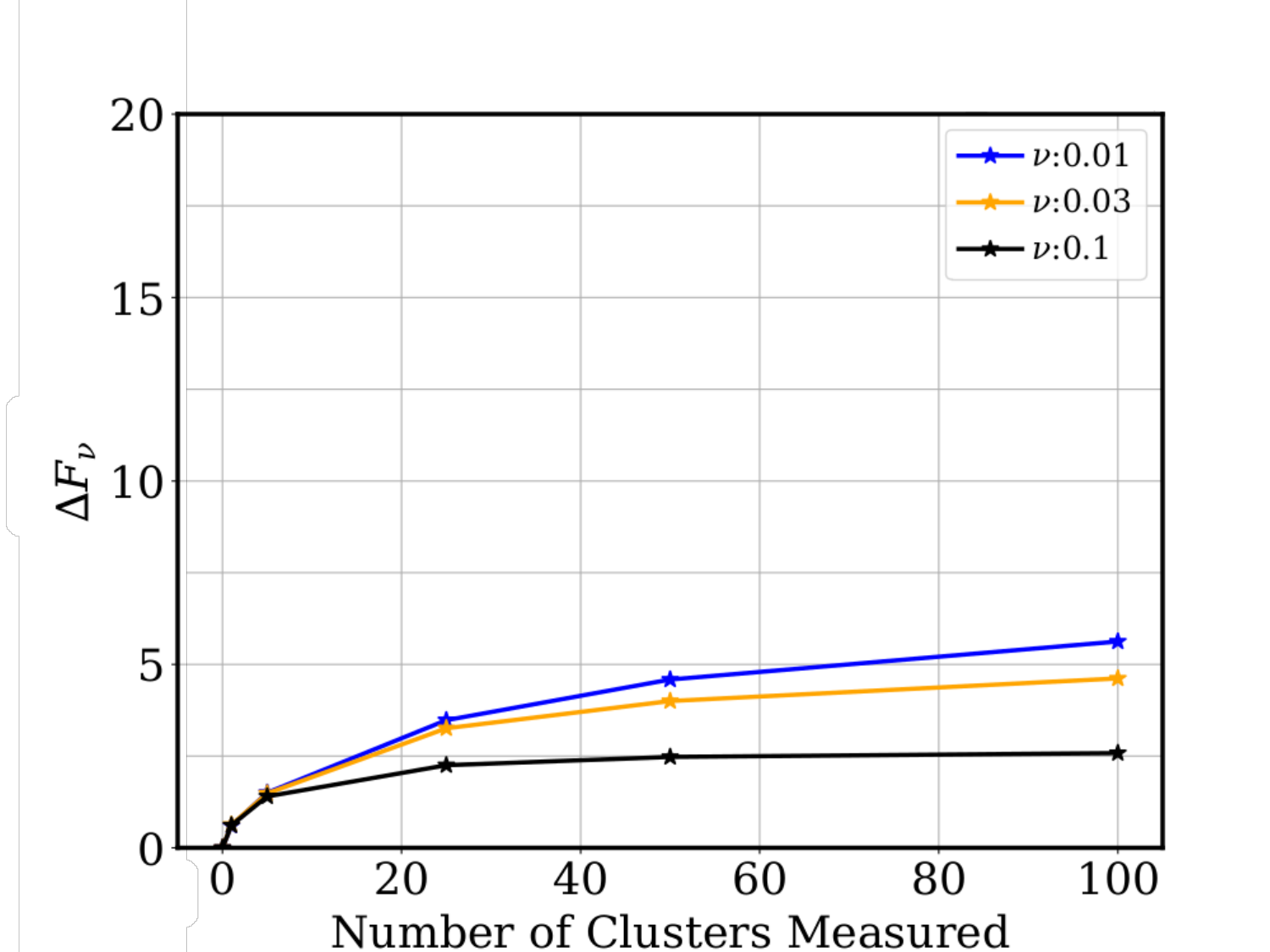}
    \end{subfigure}
    \hspace{-0.7cm}
    \begin{subfigure}{0.35\textwidth}
        \centering
        \includegraphics[width=\linewidth]{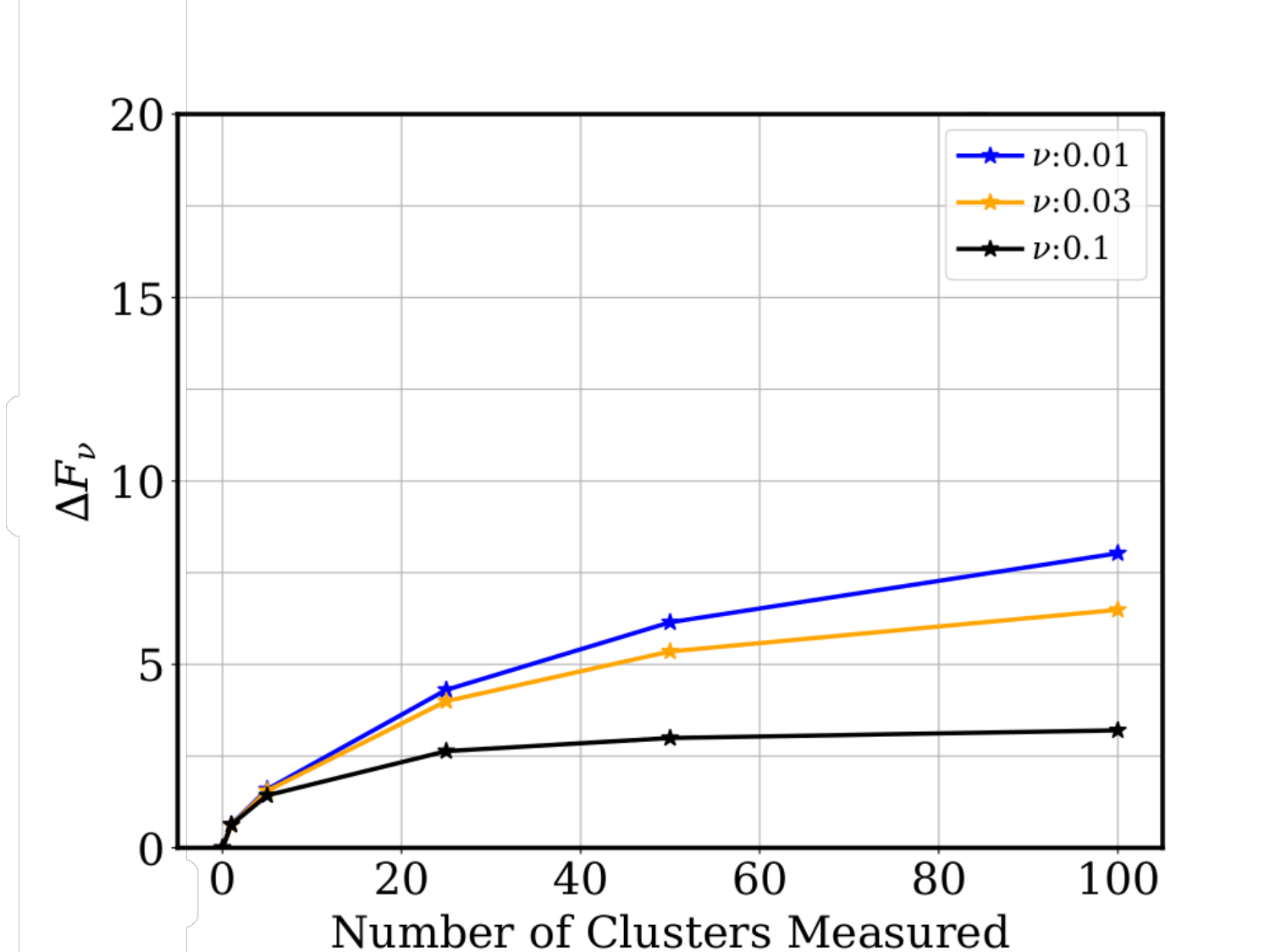}
    \end{subfigure}
    \hspace{-0.7cm}
    \begin{subfigure}{0.35\textwidth}
        \centering
        \includegraphics[width=\linewidth]{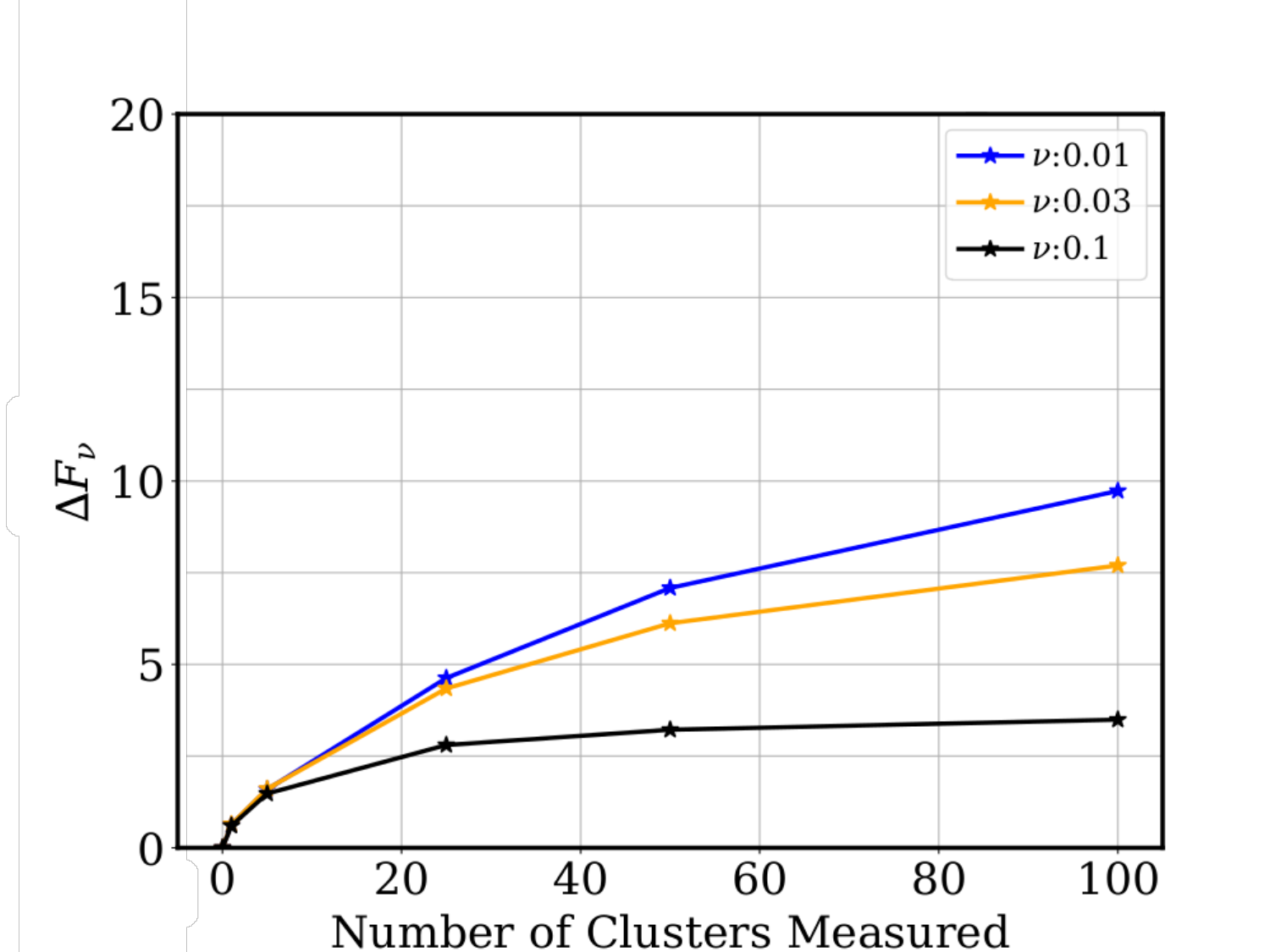}
    \end{subfigure}
    \vspace{-1cm}
    \caption{Various hypothetical surveys. The top row shows how the normalized Fisher information varies for volume-limited surveys as a function of maximum redshift. 100 clusters are measured throughout the top row. In the bottom row, $z_{\mathrm{max}}$ is kept constant at 1.0, and the number of clusters is varied. The best case out of 100 trials is shown for all scenarios. The left column corresponds to a survey covering 20\% of the sky, the middle column to a survey covering half the sky and the right column to surveys covering the whole sky. }
    \label{fig:Fn2x3}
    
\end{figure*}

Our main results are shown in Figs.~\ref{fig:FvsN} and \ref{fig:Fn2x3}, where we show the increase in the Fisher information on the cutoff parameter $k_c$ normalized by the square of the best-fit value $k_c=2.17 (H_0/c)$ gained by making the remote quadrupole measurements for a variety of survey parameters. Thus, $\Delta F$ is, colloquially speaking, the square of the ``number of $\sigma$''  improvement on $k_c$. A summary of key results is also presented in Table~\ref{tab:summary}.

For our analysis, we depict the Fisher information for a variety of noise levels, as discussed in Section~\ref{sec:signal-detec}. 

\paragraph*{Results for marginalized $\alpha$:} We first discuss the constraints on the cutoff scale $k_c$ in the more conservative case where the parameter $\alpha$, which controls the steepness of the drop in power (see the top panel in Fig.~\ref{fig:toy-model}), is marginalized. Later, we will show how these results are affected when $\alpha$ is instead fixed to the best-fit value. In Fig.~\ref{fig:FvsN}, we show the result of the gain in $F$ as the number of clusters $N$ is varied, at a given redshift, while holding the noise parameter fixed. We find that $\Delta F$ plateaus at around $N \approx 100$ across the full range of redshifts considered (depicted by the different curves in Fig.~\ref{fig:FvsN}). Of course, for a sparse sample of clusters, the value of $F$ is highly dependent on the spatial distribution of the measurements. To find the ``best'' sample (i.e. the one that maximizes $F$), we repeat the calculation many times, sampling from the $768$ pixels in a $z$-shell for each of the points at  $N<768$. For an isotropic cosmology, as considered in this work, this sample corresponds to one in which the clusters are evenly spread across the whole sky.

\begin{table}[h]
\centering
\begin{tabular}{llccc|cc}
\toprule
 & & \multicolumn{3}{c|}{$\Delta F$ ($ \alpha$ marg.)} & \multicolumn{2}{c}{$\Delta F$ ($ \alpha$ fixed)} \\
 & & $\nu=0.1$ & $0.01$ &  & $0.1$ & $0.01$ \\
\midrule
& $z_{\rm max}=1, N=100$ & 3 & 10 & & 12 & 28 \\
& $z_{\rm max}=2, N=100$ & 8 & 22 & & 25 & 39 \\
\midrule
& $z_{\rm max}=1, N=6144$ & 3 & 14 & & 12 & 40 \\
& $z_{\rm max}=2, N=10752$ & 9 & 32 & & 25 & 57 \\
\bottomrule
\end{tabular}
\caption{ A summary of results discussed in the text. The first two rows show results for sparse sky coverage, while the last two rows show results for dense coverage.  The first two columns depict results for the conservative case where $\alpha$ is marginalized (c.f. Figs.~\ref{fig:FvsN}, \ref{fig:Fn2x3}), while the last two rows depict results for the case where $\alpha$ is fixed.}
\label{tab:summary}
\end{table}

\begin{figure}
    \centering
    \includegraphics[width=9cm]{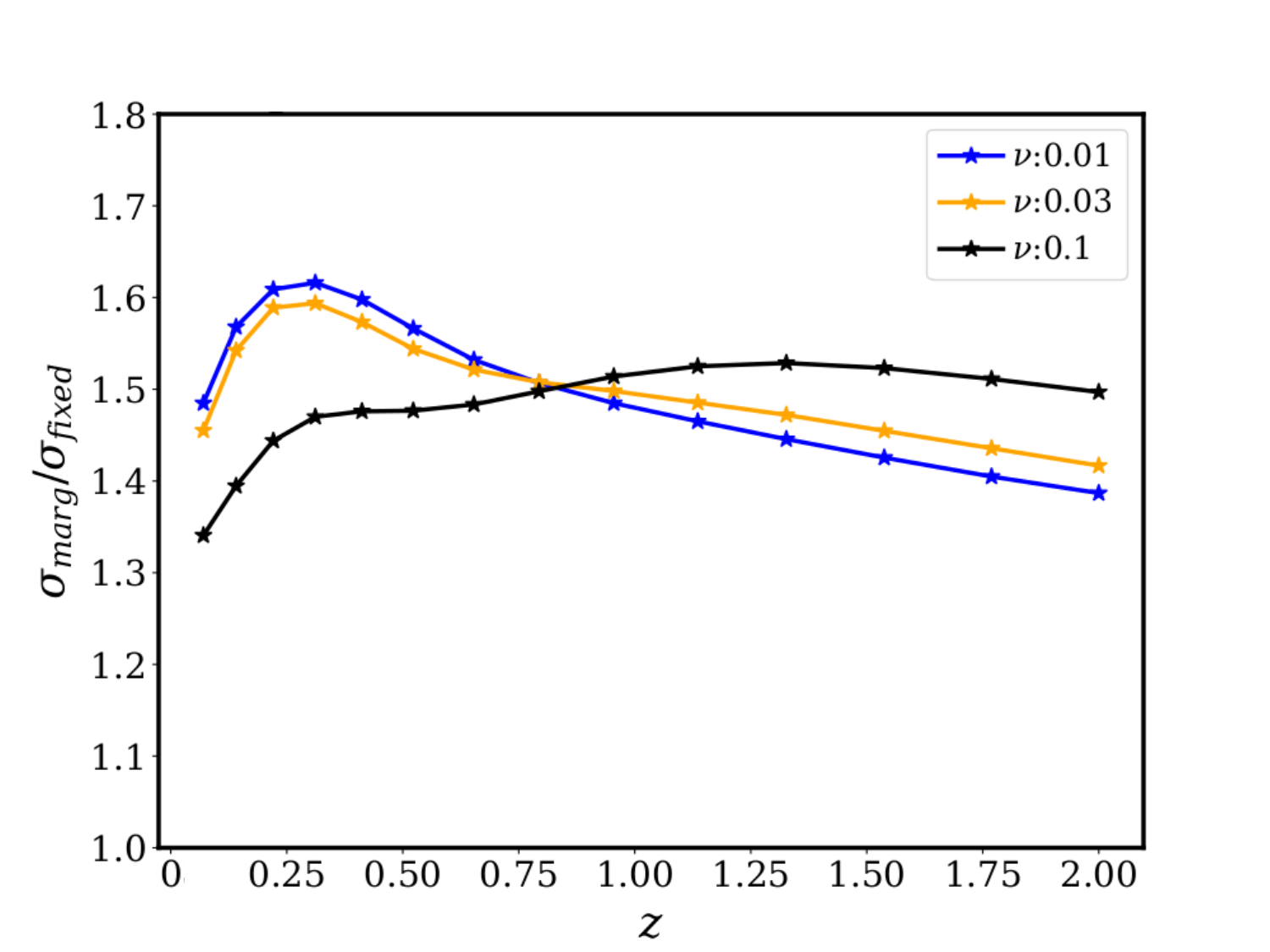}
    \caption{ We show the ratio of the uncertainty in $k_c$ achieved in the case where $\alpha$ is marginalized to the case where $\alpha$ is held fixed. The curves show the variation in this ratio with the survey depth (i.e. maximum redshift) for a variety of noise levels. Here the number of clusters is held constant at $N=100$ and $f_{\rm sky}=1.0$. }
    \label{fig:sigmaratio}
\end{figure}
In Fig.~\ref{fig:FvsN}, one can notice a significant variation in $F$ with redshift, despite two competing effects. First, one expects an increase in $F$ with increasing $z$ as the LSS probed by an observer at higher redshift is less correlated with our local CMB. On the other hand, depending on the choice of $k_c$, these observers may not probe the Fourier modes that are suppressed relative to the standard power spectrum. However, for the best-fit value of $k_c$ we consider, the quadrupole witnesses a significant suppression across the entire redshift range, as shown in the bottom panel of Fig.~\ref{fig:toy-model}. The increase in new information due to a less correlated LSS far outweighs the effects of reduced sensitivity to the suppressed Fourier modes at higher $z$, so that the largest increase in $F(z)$ comes from the highest-redshift clusters in our sample. 

In Fig.~\ref{fig:Fn2x3}, we show the effects of variations in redshift, noise levels, sky coverage, and cluster count on $F$, using the same procedure as above to find the ``best'' sample of cluster locations. In the top row of Fig.~\ref{fig:Fn2x3}, we show results for variations in $\Delta F$ in volume-limited surveys as a function of maximum redshift while the number of clusters is held fixed. For a survey that measures $100$ (well separated) clusters out to $z_{\mathrm{max}}=2$ with $20\%$ sky coverage, one can achieve $\Delta F \approx 6$ (i.e. a $20\%$ or $2.4\sigma$ improvement compared to local data alone) where the signal from each cluster is measured with a signal-to-noise ratio $1/\epsilon = 1$ (i.e. $\nu = 0.1$). This number goes up to $\Delta F \approx 10 \approx F_{\rm local}$, effectively supplying the information content of a last-scattering surface that is entirely uncorrelated with our local one, for a three-fold increase in the signal-to-noise (i.e. $\nu=0.03$). Further improvements to the noise only lead to marginal increases in $F$. 

With the \textit{number of clusters fixed} to $N=100$, increasing the sky coverage to $50\%$ leads to modest gains in the Fisher information by about $2-6$ across the range of noise levels and a further similar increase as the sky coverage is increased to $100\%$. In the case of full-sky coverage, the increase in information is limited by the number of clusters. Thus, since our simulations consist of $\approx 10^4$ voxels (see Sect.~\ref{sec:methods}), we also calculate $F$ for the hypothetical case of full and dense sky coverage. Using this entire simulated dataset, we find that $\Delta F\approx 9, 21,$ and $ 32$ for $\nu = 0.1, 0.03,$ and $ 0.01$ respectively (see Table~\ref{tab:summary}).

In the bottom panel of Fig.~\ref{fig:Fn2x3}, we show the effect of varying the number of clusters on $F$ at a fixed maximum redshift of $z_{\rm max} =1$. This is similar to Fig.~\ref{fig:FvsN}, except now the number of clusters is constant in the entire \textit{volume} up to $z_{\rm max}$, instead of on a shell at fixed $z$. In most cases of sky coverage and noise levels considered, the Fisher information has approximately plateaued by $N\approx 100$ indicating little gains to be made with further measurements. However, for the lowest noise levels in the full sky limit, the increase in $F$ is limited by the sample size. To quantify the maximum constraining power of a volume limited survey, we calculate $F$ using \textit{all} the $\approx 6000$ clusters at $z \lesssim 1$ in our simulated data and find that $\Delta F \approx 9 $ and $14$ for $\nu = 0.03$ and $0.01$ respectively. 

\paragraph*{Results for fixed $\alpha$}: We now briefly mention the effects on the results shown in Figs.~\ref{fig:FvsN} and \ref{fig:Fn2x3} for the case where $\alpha$ is held fixed. In this case, the Fisher information is simply the inverse of the relevant \textit{element} of the covariance matrix, as described in eq. (\ref{eq:fisherfixednuisance}).
We include these results because it is arguably $k_c$ that is the physically relevant parameter and not $\alpha$. For example, in an inflationary cosmology, the suppression of power on large scales may occur due to a period of kination that precedes the onset of slow-roll so that $k_c$ parameterizes the boundary between the two regimes. These results also serve to show the dependence of our Fisher analysis on model parameterizations. 

In Fig.~\ref{fig:sigmaratio} we plot the ratio of the uncertainties on $k_c$ calculated using a fixed value of $\alpha$ to the case where $\alpha$ is marginalized. In essentially all cases of sky coverage, depth, and noise, the case where $\alpha$ is fixed leads to an $\approx50\%$ decrease in the uncertainty in $k_c$. In other words, the Fisher information increases by a factor of $\approx 2.25$ when $\alpha$ is fixed. A comparison summary is presented in Table~\ref{tab:summary}. Most notable is that, with $\alpha$ fixed, the constraints on $k_c$ can be \textit{improved} by over $7\sigma$, compared to only local CMB data, with dense full-sky coverage at low signal-to-noise.

\section{Conclusions}\label{sec:conclusions}
While several models have been proposed to explain CMB anomalies, the characterization of their statistical significance often suffers from look-elsewhere effects. This effect is particularly pertinent for theories that deviate from the standard cosmology at the largest scales where cosmic variance is high. In some cases, even an improvement of $\Delta \chi^2\approx 15$ can be attributed to cosmic variance \cite{flauger2017drifting_constraints_monodromy}. In order to gauge the significance of these theories, one needs an independent dataset against which to test the hypotheses. 

In this work, we studied the information content of measuring the polarization signal of the CMB in the direction of SZ clusters, which provides a handle on when and where the CMB quadrupole is observed from. In particular, we calculated Fisher forecasts (Sec.~\ref{sec:formalism}) on a fiducial model where Fourier modes below a cutoff scale $k_c$ are exponentially suppressed compared to the standard nearly scale-invariant power spectrum of primordial fluctuations [Eq.~(\ref{eq:ps})]. We calculated observables in real space which allowed us to characterize an efficient distribution of clusters to be measured by a hypothetical remote quadrupole survey. 

Overall, our results (Sec.~\ref{sec:results}) paint an optimistic picture for improving the constraints on theories that deviate from the standard \lcdm\ cosmology in the cosmic variance regime. For the particular model we considered, one can improve the constraints on $k_c$ beyond local CMB data by $2.4\sigma$ in the most conservative case (low cluster count, signal-to-noise,  sky coverage, and redshift depth) with improvements of more than $7\sigma$ in the most optimistic case. This increase is especially apparent at higher redshifts, where the remote quadrupole signal is less correlated with the local one. 

Stage-4 CMB experiments are expected to catalog $\mathcal{O}(10^5)$ clusters, and although they will have the requisite angular resolutions ($\approx 1'$), they will not measure the signal from \textit{individual} clusters with high significance. Given the use of the remote quadrupole signal as a cosmological probe to study the nature of dark energy, CMB anomalies, and, notably, theories of quantum gravity -- which are otherwise difficult to probe -- there may be a strong case for pursuing these measurements. One challenge will be to separate the remote quadrupole signal from other confusion sources of polarization (Sec.~\ref{sec:signal-detec}). On the other hand, the other SZ polarization sources themselves carry rich astrophysical data. 

While in this work we focused on an isotropic cosmology, in future work we will extend our analysis to the feasibility of detecting a hemispheric anisotropy in the CMB, where the detailed distribution of the cluster samples is likely to play an important role in deciphering an efficient measurement scheme. 

\section*{Acknowledgements}
AA would like to thank Lloyd Knox and Andreas Albrecht for several helpful discussions.
This work was supported by Summer Research Fellowships from the University of Richmond School of Arts \& Sciences for all three authors. AA acknowledges support from the U.S. Department of Energy, Office of Science, Office of High Energy Physics QuantISED program under Contract No. KA2401032.

\bibliography{ref}

\begin{thebibliography}{45}%
\makeatletter
\providecommand \@ifxundefined [1]{%
 \@ifx{#1\undefined}
}%
\providecommand \@ifnum [1]{%
 \ifnum #1\expandafter \@firstoftwo
 \else \expandafter \@secondoftwo
 \fi
}%
\providecommand \@ifx [1]{%
 \ifx #1\expandafter \@firstoftwo
 \else \expandafter \@secondoftwo
 \fi
}%
\providecommand \natexlab [1]{#1}%
\providecommand \enquote  [1]{``#1''}%
\providecommand \bibnamefont  [1]{#1}%
\providecommand \bibfnamefont [1]{#1}%
\providecommand \citenamefont [1]{#1}%
\providecommand \href@noop [0]{\@secondoftwo}%
\providecommand \href [0]{\begingroup \@sanitize@url \@href}%
\providecommand \@href[1]{\@@startlink{#1}\@@href}%
\providecommand \@@href[1]{\endgroup#1\@@endlink}%
\providecommand \@sanitize@url [0]{\catcode `\\12\catcode `\$12\catcode `\&12\catcode `\#12\catcode `\^12\catcode `\_12\catcode `\%12\relax}%
\providecommand \@@startlink[1]{}%
\providecommand \@@endlink[0]{}%
\providecommand \url  [0]{\begingroup\@sanitize@url \@url }%
\providecommand \@url [1]{\endgroup\@href {#1}{\urlprefix }}%
\providecommand \urlprefix  [0]{URL }%
\providecommand \Eprint [0]{\href }%
\providecommand \doibase [0]{https://doi.org/}%
\providecommand \selectlanguage [0]{\@gobble}%
\providecommand \bibinfo  [0]{\@secondoftwo}%
\providecommand \bibfield  [0]{\@secondoftwo}%
\providecommand \translation [1]{[#1]}%
\providecommand \BibitemOpen [0]{}%
\providecommand \bibitemStop [0]{}%
\providecommand \bibitemNoStop [0]{.\EOS\space}%
\providecommand \EOS [0]{\spacefactor3000\relax}%
\providecommand \BibitemShut  [1]{\csname bibitem#1\endcsname}%
\let\auto@bib@innerbib\@empty
\bibitem [{\citenamefont {{Smoot}}\ \emph {et~al.}(1992)\citenamefont {{Smoot}}, \citenamefont {{Bennett}}, \citenamefont {{Kogut}}, \citenamefont {{Wright}} \emph {et~al.}}]{cobe}%
  \BibitemOpen
  \bibfield  {author} {\bibinfo {author} {\bibfnamefont {G.~F.}\ \bibnamefont {{Smoot}}}, \bibinfo {author} {\bibfnamefont {C.~L.}\ \bibnamefont {{Bennett}}}, \bibinfo {author} {\bibfnamefont {A.}~\bibnamefont {{Kogut}}}, \bibinfo {author} {\bibnamefont {{Wright}}}, \emph {et~al.},\ }\bibfield  {title} {\bibinfo {title} {{Structure in the COBE Differential Microwave Radiometer First-Year Maps}},\ }\href {https://doi.org/10.1086/186504} {\bibfield  {journal} {\bibinfo  {journal} {\apjl}\ }\textbf {\bibinfo {volume} {396}},\ \bibinfo {pages} {L1} (\bibinfo {year} {1992})}\BibitemShut {NoStop}%
\bibitem [{\citenamefont {Bennett}\ \emph {et~al.}(2013)\citenamefont {Bennett}, \citenamefont {Larson}, \citenamefont {Weiland} \emph {et~al.}}]{wmapresults}%
  \BibitemOpen
  \bibfield  {author} {\bibinfo {author} {\bibfnamefont {C.~L.}\ \bibnamefont {Bennett}}, \bibinfo {author} {\bibfnamefont {D.}~\bibnamefont {Larson}}, \bibinfo {author} {\bibfnamefont {J.~L.}\ \bibnamefont {Weiland}}, \emph {et~al.},\ }\bibfield  {title} {\bibinfo {title} {Nine-year wilkinson microwave anisotropy probe ( wmap ) observations: Final maps and results},\ }\href {https://doi.org/10.1088/0067-0049/208/2/20} {\bibfield  {journal} {\bibinfo  {journal} {The Astrophysical Journal Supplement Series}\ }\textbf {\bibinfo {volume} {208}},\ \bibinfo {pages} {20} (\bibinfo {year} {2013})}\BibitemShut {NoStop}%
\bibitem [{\citenamefont {{Planck Collaboration}}\ \emph {et~al.}(2020)\citenamefont {{Planck Collaboration}}, \citenamefont {{Aghanim}} \emph {et~al.}}]{planckresults}%
  \BibitemOpen
  \bibfield  {author} {\bibinfo {author} {\bibnamefont {{Planck Collaboration}}}, \bibinfo {author} {\bibfnamefont {N.}~\bibnamefont {{Aghanim}}}, \emph {et~al.},\ }\bibfield  {title} {\bibinfo {title} {{Planck 2018 results. I. Overview and the cosmological legacy of Planck}},\ }\href {https://doi.org/10.1051/0004-6361/201833880} {\bibfield  {journal} {\bibinfo  {journal} {\aap}\ }\textbf {\bibinfo {volume} {641}},\ \bibinfo {eid} {A1} (\bibinfo {year} {2020})},\ \Eprint {https://arxiv.org/abs/1807.06205} {arXiv:1807.06205 [astro-ph.CO]} \BibitemShut {NoStop}%
\bibitem [{\citenamefont {{South Pole Telescope Collaboration}}\ \emph {et~al.}(2018)\citenamefont {{South Pole Telescope Collaboration}}, \citenamefont {{Chown}}, \citenamefont {{Omori}}, \citenamefont {{Aylor}} \emph {et~al.}}]{spt}%
  \BibitemOpen
  \bibfield  {author} {\bibinfo {author} {\bibnamefont {{South Pole Telescope Collaboration}}}, \bibinfo {author} {\bibfnamefont {R.}~\bibnamefont {{Chown}}}, \bibinfo {author} {\bibfnamefont {Y.}~\bibnamefont {{Omori}}}, \bibinfo {author} {\bibfnamefont {K.}~\bibnamefont {{Aylor}}}, \emph {et~al.},\ }\bibfield  {title} {\bibinfo {title} {{Maps of the Southern Millimeter-wave Sky from Combined 2500 deg$^{2}$ SPT-SZ and Planck Temperature Data}},\ }\href {https://doi.org/10.3847/1538-4365/aae694} {\bibfield  {journal} {\bibinfo  {journal} {\apjs}\ }\textbf {\bibinfo {volume} {239}},\ \bibinfo {eid} {10} (\bibinfo {year} {2018})},\ \Eprint {https://arxiv.org/abs/1803.10682} {arXiv:1803.10682 [astro-ph.CO]} \BibitemShut {NoStop}%
\bibitem [{\citenamefont {{Ade}}\ \emph {et~al.}(2022)\citenamefont {{Ade}}, \citenamefont {{Ahmed}}, \citenamefont {{Amiri}} \emph {et~al.}}]{bicep}%
  \BibitemOpen
  \bibfield  {author} {\bibinfo {author} {\bibfnamefont {P.~A.~R.}\ \bibnamefont {{Ade}}}, \bibinfo {author} {\bibfnamefont {Z.}~\bibnamefont {{Ahmed}}}, \bibinfo {author} {\bibfnamefont {M.}~\bibnamefont {{Amiri}}}, \emph {et~al.},\ }\bibfield  {title} {\bibinfo {title} {{BICEP/Keck XV: The BICEP3 Cosmic Microwave Background Polarimeter and the First Three-year Data Set}},\ }\href {https://doi.org/10.3847/1538-4357/ac4886} {\bibfield  {journal} {\bibinfo  {journal} {\apj}\ }\textbf {\bibinfo {volume} {927}},\ \bibinfo {eid} {77} (\bibinfo {year} {2022})},\ \Eprint {https://arxiv.org/abs/2110.00482} {arXiv:2110.00482 [astro-ph.IM]} \BibitemShut {NoStop}%
\bibitem [{\citenamefont {Schwarz}\ \emph {et~al.}(2016)\citenamefont {Schwarz}, \citenamefont {Copi}, \citenamefont {Huterer},\ and\ \citenamefont {Starkman}}]{Schwarz_2016_CMBanomalies}%
  \BibitemOpen
  \bibfield  {author} {\bibinfo {author} {\bibfnamefont {D.~J.}\ \bibnamefont {Schwarz}}, \bibinfo {author} {\bibfnamefont {C.~J.}\ \bibnamefont {Copi}}, \bibinfo {author} {\bibfnamefont {D.}~\bibnamefont {Huterer}},\ and\ \bibinfo {author} {\bibfnamefont {G.~D.}\ \bibnamefont {Starkman}},\ }\bibfield  {title} {\bibinfo {title} {Cmb anomalies after planck},\ }\href {https://doi.org/10.1088/0264-9381/33/18/184001} {\bibfield  {journal} {\bibinfo  {journal} {Classical and Quantum Gravity}\ }\textbf {\bibinfo {volume} {33}},\ \bibinfo {pages} {184001} (\bibinfo {year} {2016})}\BibitemShut {NoStop}%
\bibitem [{\citenamefont {Bennett}\ \emph {et~al.}(2011)\citenamefont {Bennett}, \citenamefont {Hill}, \citenamefont {Hinshaw}, \citenamefont {Larson}, \citenamefont {Smith}, \citenamefont {Dunkley}, \citenamefont {Gold}, \citenamefont {Halpern}, \citenamefont {Jarosik}, \citenamefont {Kogut} \emph {et~al.}}]{bennett2011seven_cmbanomalies}%
  \BibitemOpen
  \bibfield  {author} {\bibinfo {author} {\bibfnamefont {C.}~\bibnamefont {Bennett}}, \bibinfo {author} {\bibfnamefont {R.}~\bibnamefont {Hill}}, \bibinfo {author} {\bibfnamefont {G.}~\bibnamefont {Hinshaw}}, \bibinfo {author} {\bibfnamefont {D.}~\bibnamefont {Larson}}, \bibinfo {author} {\bibfnamefont {K.}~\bibnamefont {Smith}}, \bibinfo {author} {\bibfnamefont {J.}~\bibnamefont {Dunkley}}, \bibinfo {author} {\bibfnamefont {B.}~\bibnamefont {Gold}}, \bibinfo {author} {\bibfnamefont {M.}~\bibnamefont {Halpern}}, \bibinfo {author} {\bibfnamefont {N.}~\bibnamefont {Jarosik}}, \bibinfo {author} {\bibfnamefont {A.}~\bibnamefont {Kogut}}, \emph {et~al.},\ }\bibfield  {title} {\bibinfo {title} {Seven-year wilkinson microwave anisotropy probe (wmap*) observations: Are there cosmic microwave background anomalies?},\ }\href@noop {} {\bibfield  {journal} {\bibinfo  {journal} {The Astrophysical journal supplement series}\ }\textbf {\bibinfo {volume} {192}},\ \bibinfo {pages} {17} (\bibinfo {year} {2011})}\BibitemShut
  {NoStop}%
\bibitem [{\citenamefont {Silverstein}\ and\ \citenamefont {Westphal}(2008)}]{silverstein2008monodromy}%
  \BibitemOpen
  \bibfield  {author} {\bibinfo {author} {\bibfnamefont {E.}~\bibnamefont {Silverstein}}\ and\ \bibinfo {author} {\bibfnamefont {A.}~\bibnamefont {Westphal}},\ }\bibfield  {title} {\bibinfo {title} {Monodromy in the cmb: gravity waves and string inflation},\ }\href@noop {} {\bibfield  {journal} {\bibinfo  {journal} {Physical Review D—Particles, Fields, Gravitation, and Cosmology}\ }\textbf {\bibinfo {volume} {78}},\ \bibinfo {pages} {106003} (\bibinfo {year} {2008})}\BibitemShut {NoStop}%
\bibitem [{\citenamefont {McAllister}\ \emph {et~al.}(2010)\citenamefont {McAllister}, \citenamefont {Silverstein},\ and\ \citenamefont {Westphal}}]{mcallister2010gravity_monodromy}%
  \BibitemOpen
  \bibfield  {author} {\bibinfo {author} {\bibfnamefont {L.}~\bibnamefont {McAllister}}, \bibinfo {author} {\bibfnamefont {E.}~\bibnamefont {Silverstein}},\ and\ \bibinfo {author} {\bibfnamefont {A.}~\bibnamefont {Westphal}},\ }\bibfield  {title} {\bibinfo {title} {Gravity waves and linear inflation from axion monodromy},\ }\href@noop {} {\bibfield  {journal} {\bibinfo  {journal} {Physical Review D—Particles, Fields, Gravitation, and Cosmology}\ }\textbf {\bibinfo {volume} {82}},\ \bibinfo {pages} {046003} (\bibinfo {year} {2010})}\BibitemShut {NoStop}%
\bibitem [{\citenamefont {Kaloper}\ \emph {et~al.}(2011)\citenamefont {Kaloper}, \citenamefont {Lawrence},\ and\ \citenamefont {Sorbo}}]{kaloper2011ignoble}%
  \BibitemOpen
  \bibfield  {author} {\bibinfo {author} {\bibfnamefont {N.}~\bibnamefont {Kaloper}}, \bibinfo {author} {\bibfnamefont {A.}~\bibnamefont {Lawrence}},\ and\ \bibinfo {author} {\bibfnamefont {L.}~\bibnamefont {Sorbo}},\ }\bibfield  {title} {\bibinfo {title} {An ignoble approach to large field inflation},\ }\href@noop {} {\bibfield  {journal} {\bibinfo  {journal} {Journal of Cosmology and Astroparticle Physics}\ }\textbf {\bibinfo {volume} {2011}}\bibinfo  {number} { (03)},\ \bibinfo {pages} {023}}\BibitemShut {NoStop}%
\bibitem [{\citenamefont {Adil}\ \emph {et~al.}(2023)\citenamefont {Adil}, \citenamefont {Albrecht}, \citenamefont {Baunach}, \citenamefont {Holman}, \citenamefont {Ribeiro},\ and\ \citenamefont {Richard}}]{adil2023entanglement}%
  \BibitemOpen
\bibfield  {number} {  }\bibfield  {author} {\bibinfo {author} {\bibfnamefont {A.}~\bibnamefont {Adil}}, \bibinfo {author} {\bibfnamefont {A.}~\bibnamefont {Albrecht}}, \bibinfo {author} {\bibfnamefont {R.}~\bibnamefont {Baunach}}, \bibinfo {author} {\bibfnamefont {R.}~\bibnamefont {Holman}}, \bibinfo {author} {\bibfnamefont {R.~H.}\ \bibnamefont {Ribeiro}},\ and\ \bibinfo {author} {\bibfnamefont {B.~J.}\ \bibnamefont {Richard}},\ }\bibfield  {title} {\bibinfo {title} {Entanglement masquerading in the cmb},\ }\href@noop {} {\bibfield  {journal} {\bibinfo  {journal} {Journal of Cosmology and Astroparticle Physics}\ }\textbf {\bibinfo {volume} {2023}}\bibinfo  {number} { (06)},\ \bibinfo {pages} {024}}\BibitemShut {NoStop}%
\bibitem [{\citenamefont {Chluba}\ \emph {et~al.}(2015)\citenamefont {Chluba}, \citenamefont {Hamann},\ and\ \citenamefont {Patil}}]{Chluba_2015_subodhpatel}%
  \BibitemOpen
\bibfield  {number} {  }\bibfield  {author} {\bibinfo {author} {\bibfnamefont {J.}~\bibnamefont {Chluba}}, \bibinfo {author} {\bibfnamefont {J.}~\bibnamefont {Hamann}},\ and\ \bibinfo {author} {\bibfnamefont {S.~P.}\ \bibnamefont {Patil}},\ }\bibfield  {title} {\bibinfo {title} {Features and new physical scales in primordial observables: Theory and observation},\ }\href {https://doi.org/10.1142/s0218271815300232} {\bibfield  {journal} {\bibinfo  {journal} {International Journal of Modern Physics D}\ }\textbf {\bibinfo {volume} {24}},\ \bibinfo {pages} {1530023} (\bibinfo {year} {2015})}\BibitemShut {NoStop}%
\bibitem [{\citenamefont {Akrami}\ \emph {et~al.}(2020)\citenamefont {Akrami}, \citenamefont {Arroja}, \citenamefont {Ashdown}, \citenamefont {Aumont}, \citenamefont {Baccigalupi}, \citenamefont {Ballardini}, \citenamefont {Banday}, \citenamefont {Barreiro}, \citenamefont {Bartolo}, \citenamefont {Basak} \emph {et~al.}}]{akrami2020planck}%
  \BibitemOpen
  \bibfield  {author} {\bibinfo {author} {\bibfnamefont {Y.}~\bibnamefont {Akrami}}, \bibinfo {author} {\bibfnamefont {F.}~\bibnamefont {Arroja}}, \bibinfo {author} {\bibfnamefont {M.}~\bibnamefont {Ashdown}}, \bibinfo {author} {\bibfnamefont {J.}~\bibnamefont {Aumont}}, \bibinfo {author} {\bibfnamefont {C.}~\bibnamefont {Baccigalupi}}, \bibinfo {author} {\bibfnamefont {M.}~\bibnamefont {Ballardini}}, \bibinfo {author} {\bibfnamefont {A.~J.}\ \bibnamefont {Banday}}, \bibinfo {author} {\bibfnamefont {R.}~\bibnamefont {Barreiro}}, \bibinfo {author} {\bibfnamefont {N.}~\bibnamefont {Bartolo}}, \bibinfo {author} {\bibfnamefont {S.}~\bibnamefont {Basak}}, \emph {et~al.},\ }\bibfield  {title} {\bibinfo {title} {Planck 2018 results-x. constraints on inflation},\ }\href@noop {} {\bibfield  {journal} {\bibinfo  {journal} {Astronomy \& Astrophysics}\ }\textbf {\bibinfo {volume} {641}},\ \bibinfo {pages} {A10} (\bibinfo {year} {2020})}\BibitemShut {NoStop}%
\bibitem [{\citenamefont {Kamionkowski}\ and\ \citenamefont {Loeb}(1997)}]{kamionkowski1997getting}%
  \BibitemOpen
  \bibfield  {author} {\bibinfo {author} {\bibfnamefont {M.}~\bibnamefont {Kamionkowski}}\ and\ \bibinfo {author} {\bibfnamefont {A.}~\bibnamefont {Loeb}},\ }\bibfield  {title} {\bibinfo {title} {Getting around cosmic variance},\ }\href@noop {} {\bibfield  {journal} {\bibinfo  {journal} {Physical Review D}\ }\textbf {\bibinfo {volume} {56}},\ \bibinfo {pages} {4511} (\bibinfo {year} {1997})}\BibitemShut {NoStop}%
\bibitem [{\citenamefont {Sazonov}\ and\ \citenamefont {Sunyaev}(1999)}]{sazonov1999microwave}%
  \BibitemOpen
  \bibfield  {author} {\bibinfo {author} {\bibfnamefont {S.}~\bibnamefont {Sazonov}}\ and\ \bibinfo {author} {\bibfnamefont {R.}~\bibnamefont {Sunyaev}},\ }\bibfield  {title} {\bibinfo {title} {Microwave polarization in the direction of galaxy clusters induced by the cmb quadrupole anisotropy},\ }\href@noop {} {\bibfield  {journal} {\bibinfo  {journal} {Monthly Notices of the Royal Astronomical Society}\ }\textbf {\bibinfo {volume} {310}},\ \bibinfo {pages} {765} (\bibinfo {year} {1999})}\BibitemShut {NoStop}%
\bibitem [{\citenamefont {Sunyaev}\ and\ \citenamefont {Zeldovich}(1980)}]{sunyaev1980velocity}%
  \BibitemOpen
  \bibfield  {author} {\bibinfo {author} {\bibfnamefont {R.}~\bibnamefont {Sunyaev}}\ and\ \bibinfo {author} {\bibfnamefont {Y.~B.}\ \bibnamefont {Zeldovich}},\ }\bibfield  {title} {\bibinfo {title} {The velocity of clusters of galaxies relative to the microwave background-the possibility of its measurement},\ }\href@noop {} {\bibfield  {journal} {\bibinfo  {journal} {Monthly Notices of the Royal Astronomical Society, vol. 190, Feb. 1980, p. 413-420.}\ }\textbf {\bibinfo {volume} {190}},\ \bibinfo {pages} {413} (\bibinfo {year} {1980})}\BibitemShut {NoStop}%
\bibitem [{\citenamefont {Sunyaev}\ and\ \citenamefont {Zel'dovich}(1981)}]{syunyaev}%
  \BibitemOpen
  \bibfield  {author} {\bibinfo {author} {\bibfnamefont {R.~A.}\ \bibnamefont {Sunyaev}}\ and\ \bibinfo {author} {\bibfnamefont {Y.~B.}\ \bibnamefont {Zel'dovich}},\ }\href@noop {} {\emph {\bibinfo {title} {Astrophysics and Space Physics Reviews, Section E}}},\ Vol.~\bibinfo {volume} {1}\ (\bibinfo  {publisher} {Harwood Academic Publisher},\ \bibinfo {year} {1981})\ pp.\ \bibinfo {pages} {1--60}\BibitemShut {NoStop}%
\bibitem [{\citenamefont {Bunn}(2006)}]{Bunn2006probing}%
  \BibitemOpen
  \bibfield  {author} {\bibinfo {author} {\bibfnamefont {E.~F.}\ \bibnamefont {Bunn}},\ }\bibfield  {title} {\bibinfo {title} {Probing the universe on gigaparsec scales with remote cosmic microwave background quadrupole measurements},\ }\bibfield  {journal} {\bibinfo  {journal} {Physical Review D}\ }\textbf {\bibinfo {volume} {73}},\ \href {https://doi.org/10.1103/physrevd.73.123517} {10.1103/physrevd.73.123517} (\bibinfo {year} {2006})\BibitemShut {NoStop}%
\bibitem [{\citenamefont {Hall}\ and\ \citenamefont {Challinor}(2014)}]{hall2014detecting_challinor}%
  \BibitemOpen
  \bibfield  {author} {\bibinfo {author} {\bibfnamefont {A.}~\bibnamefont {Hall}}\ and\ \bibinfo {author} {\bibfnamefont {A.}~\bibnamefont {Challinor}},\ }\bibfield  {title} {\bibinfo {title} {Detecting the polarization induced by scattering of the microwave background quadrupole in galaxy clusters},\ }\href@noop {} {\bibfield  {journal} {\bibinfo  {journal} {Physical Review D}\ }\textbf {\bibinfo {volume} {90}},\ \bibinfo {pages} {063518} (\bibinfo {year} {2014})}\BibitemShut {NoStop}%
\bibitem [{\citenamefont {Portsmouth}(2004)}]{portsmouth2004analysis}%
  \BibitemOpen
  \bibfield  {author} {\bibinfo {author} {\bibfnamefont {J.}~\bibnamefont {Portsmouth}},\ }\bibfield  {title} {\bibinfo {title} {Analysis of the kamionkowski-loeb method of reducing cosmic variance with cmb polarization},\ }\href@noop {} {\bibfield  {journal} {\bibinfo  {journal} {Physical Review D—Particles, Fields, Gravitation, and Cosmology}\ }\textbf {\bibinfo {volume} {70}},\ \bibinfo {pages} {063504} (\bibinfo {year} {2004})}\BibitemShut {NoStop}%
\bibitem [{\citenamefont {Seto}\ and\ \citenamefont {Pierpaoli}(2005)}]{seto2005probing}%
  \BibitemOpen
  \bibfield  {author} {\bibinfo {author} {\bibfnamefont {N.}~\bibnamefont {Seto}}\ and\ \bibinfo {author} {\bibfnamefont {E.}~\bibnamefont {Pierpaoli}},\ }\bibfield  {title} {\bibinfo {title} {Probing the largest scale structure in the universe with polarization map of galaxy clusters},\ }\href@noop {} {\bibfield  {journal} {\bibinfo  {journal} {Physical Review Letters}\ }\textbf {\bibinfo {volume} {95}},\ \bibinfo {pages} {101302} (\bibinfo {year} {2005})}\BibitemShut {NoStop}%
\bibitem [{\citenamefont {Louis}\ \emph {et~al.}(2017)\citenamefont {Louis}, \citenamefont {Bunn}, \citenamefont {Wandelt},\ and\ \citenamefont {Silk}}]{louis2017measuring_bunn}%
  \BibitemOpen
  \bibfield  {author} {\bibinfo {author} {\bibfnamefont {T.}~\bibnamefont {Louis}}, \bibinfo {author} {\bibfnamefont {E.~F.}\ \bibnamefont {Bunn}}, \bibinfo {author} {\bibfnamefont {B.}~\bibnamefont {Wandelt}},\ and\ \bibinfo {author} {\bibfnamefont {J.}~\bibnamefont {Silk}},\ }\bibfield  {title} {\bibinfo {title} {Measuring polarized emission in clusters in the cmb s4 era},\ }\href@noop {} {\bibfield  {journal} {\bibinfo  {journal} {Physical Review D}\ }\textbf {\bibinfo {volume} {96}},\ \bibinfo {pages} {123509} (\bibinfo {year} {2017})}\BibitemShut {NoStop}%
\bibitem [{\citenamefont {Meyers}\ \emph {et~al.}(2018)\citenamefont {Meyers}, \citenamefont {Meerburg}, \citenamefont {Van~Engelen},\ and\ \citenamefont {Battaglia}}]{meyers2018beyond_reionization}%
  \BibitemOpen
  \bibfield  {author} {\bibinfo {author} {\bibfnamefont {J.}~\bibnamefont {Meyers}}, \bibinfo {author} {\bibfnamefont {P.~D.}\ \bibnamefont {Meerburg}}, \bibinfo {author} {\bibfnamefont {A.}~\bibnamefont {Van~Engelen}},\ and\ \bibinfo {author} {\bibfnamefont {N.}~\bibnamefont {Battaglia}},\ }\bibfield  {title} {\bibinfo {title} {Beyond cmb cosmic variance limits on reionization with the polarized sunyaev-zel’dovich effect},\ }\href@noop {} {\bibfield  {journal} {\bibinfo  {journal} {Physical Review D}\ }\textbf {\bibinfo {volume} {97}},\ \bibinfo {pages} {103505} (\bibinfo {year} {2018})}\BibitemShut {NoStop}%
\bibitem [{\citenamefont {Lee}\ \emph {et~al.}(2022)\citenamefont {Lee}, \citenamefont {Hotinli},\ and\ \citenamefont {Kamionkowski}}]{lee2022probing_birefringence}%
  \BibitemOpen
  \bibfield  {author} {\bibinfo {author} {\bibfnamefont {N.}~\bibnamefont {Lee}}, \bibinfo {author} {\bibfnamefont {S.~C.}\ \bibnamefont {Hotinli}},\ and\ \bibinfo {author} {\bibfnamefont {M.}~\bibnamefont {Kamionkowski}},\ }\bibfield  {title} {\bibinfo {title} {Probing cosmic birefringence with polarized sunyaev-zel’dovich tomography},\ }\href@noop {} {\bibfield  {journal} {\bibinfo  {journal} {Physical Review D}\ }\textbf {\bibinfo {volume} {106}},\ \bibinfo {pages} {083518} (\bibinfo {year} {2022})}\BibitemShut {NoStop}%
\bibitem [{\citenamefont {Namikawa}\ and\ \citenamefont {Obata}(2023)}]{namikawa2023cosmic_birefringence}%
  \BibitemOpen
  \bibfield  {author} {\bibinfo {author} {\bibfnamefont {T.}~\bibnamefont {Namikawa}}\ and\ \bibinfo {author} {\bibfnamefont {I.}~\bibnamefont {Obata}},\ }\bibfield  {title} {\bibinfo {title} {Cosmic birefringence tomography with polarized sunyaev-zel’dovich effect},\ }\href@noop {} {\bibfield  {journal} {\bibinfo  {journal} {Physical Review D}\ }\textbf {\bibinfo {volume} {108}},\ \bibinfo {pages} {083510} (\bibinfo {year} {2023})}\BibitemShut {NoStop}%
\bibitem [{\citenamefont {Pan}\ and\ \citenamefont {Johnson}(2019)}]{pan2019forecasted_modified_gravity}%
  \BibitemOpen
  \bibfield  {author} {\bibinfo {author} {\bibfnamefont {Z.}~\bibnamefont {Pan}}\ and\ \bibinfo {author} {\bibfnamefont {M.~C.}\ \bibnamefont {Johnson}},\ }\bibfield  {title} {\bibinfo {title} {Forecasted constraints on modified gravity from sunyaev-zel’dovich tomography},\ }\href@noop {} {\bibfield  {journal} {\bibinfo  {journal} {Physical Review D}\ }\textbf {\bibinfo {volume} {100}},\ \bibinfo {pages} {083522} (\bibinfo {year} {2019})}\BibitemShut {NoStop}%
\bibitem [{\citenamefont {Deutsch}\ \emph {et~al.}(2018{\natexlab{a}})\citenamefont {Deutsch}, \citenamefont {Johnson}, \citenamefont {M{\"u}nchmeyer},\ and\ \citenamefont {Terrana}}]{deutsch2018polarized_hemispheric_anisotropy}%
  \BibitemOpen
  \bibfield  {author} {\bibinfo {author} {\bibfnamefont {A.-S.}\ \bibnamefont {Deutsch}}, \bibinfo {author} {\bibfnamefont {M.~C.}\ \bibnamefont {Johnson}}, \bibinfo {author} {\bibfnamefont {M.}~\bibnamefont {M{\"u}nchmeyer}},\ and\ \bibinfo {author} {\bibfnamefont {A.}~\bibnamefont {Terrana}},\ }\bibfield  {title} {\bibinfo {title} {Polarized sunyaev zel'dovich tomography},\ }\href@noop {} {\bibfield  {journal} {\bibinfo  {journal} {Journal of Cosmology and Astroparticle Physics}\ }\textbf {\bibinfo {volume} {2018}}\bibinfo  {number} { (04)},\ \bibinfo {pages} {034}}\BibitemShut {NoStop}%
\bibitem [{\citenamefont {Deutsch}\ \emph {et~al.}(2018{\natexlab{b}})\citenamefont {Deutsch}, \citenamefont {Dimastrogiovanni}, \citenamefont {Johnson}, \citenamefont {M{\"u}nchmeyer},\ and\ \citenamefont {Terrana}}]{deutsch2018reconstruction}%
  \BibitemOpen
\bibfield  {number} {  }\bibfield  {author} {\bibinfo {author} {\bibfnamefont {A.-S.}\ \bibnamefont {Deutsch}}, \bibinfo {author} {\bibfnamefont {E.}~\bibnamefont {Dimastrogiovanni}}, \bibinfo {author} {\bibfnamefont {M.~C.}\ \bibnamefont {Johnson}}, \bibinfo {author} {\bibfnamefont {M.}~\bibnamefont {M{\"u}nchmeyer}},\ and\ \bibinfo {author} {\bibfnamefont {A.}~\bibnamefont {Terrana}},\ }\bibfield  {title} {\bibinfo {title} {Reconstruction of the remote dipole and quadrupole fields from the kinetic sunyaev zel’dovich and polarized sunyaev zel’dovich effects},\ }\href@noop {} {\bibfield  {journal} {\bibinfo  {journal} {Physical Review D}\ }\textbf {\bibinfo {volume} {98}},\ \bibinfo {pages} {123501} (\bibinfo {year} {2018}{\natexlab{b}})}\BibitemShut {NoStop}%
\bibitem [{\citenamefont {Cayuso}\ and\ \citenamefont {Johnson}(2020)}]{cayuso2020towards}%
  \BibitemOpen
  \bibfield  {author} {\bibinfo {author} {\bibfnamefont {J.~I.}\ \bibnamefont {Cayuso}}\ and\ \bibinfo {author} {\bibfnamefont {M.~C.}\ \bibnamefont {Johnson}},\ }\bibfield  {title} {\bibinfo {title} {Towards testing cmb anomalies using the kinetic and polarized sunyaev-zel’dovich effects},\ }\href@noop {} {\bibfield  {journal} {\bibinfo  {journal} {Physical Review D}\ }\textbf {\bibinfo {volume} {101}},\ \bibinfo {pages} {123508} (\bibinfo {year} {2020})}\BibitemShut {NoStop}%
\bibitem [{\citenamefont {Contaldi}\ \emph {et~al.}(2003)\citenamefont {Contaldi}, \citenamefont {Peloso}, \citenamefont {Kofman},\ and\ \citenamefont {Linde}}]{contaldi2003suppressing}%
  \BibitemOpen
  \bibfield  {author} {\bibinfo {author} {\bibfnamefont {C.~R.}\ \bibnamefont {Contaldi}}, \bibinfo {author} {\bibfnamefont {M.}~\bibnamefont {Peloso}}, \bibinfo {author} {\bibfnamefont {L.}~\bibnamefont {Kofman}},\ and\ \bibinfo {author} {\bibfnamefont {A.}~\bibnamefont {Linde}},\ }\bibfield  {title} {\bibinfo {title} {Suppressing the lower multipoles in the cmb anisotropies},\ }\href@noop {} {\bibfield  {journal} {\bibinfo  {journal} {Journal of Cosmology and Astroparticle Physics}\ }\textbf {\bibinfo {volume} {2003}}\bibinfo  {number} { (07)},\ \bibinfo {pages} {002}}\BibitemShut {NoStop}%
\bibitem [{\citenamefont {Padmanabhan}(2003)}]{padmanabhan2003cosmological}%
  \BibitemOpen
\bibfield  {number} {  }\bibfield  {author} {\bibinfo {author} {\bibfnamefont {T.}~\bibnamefont {Padmanabhan}},\ }\bibfield  {title} {\bibinfo {title} {Cosmological constant—the weight of the vacuum},\ }\href@noop {} {\bibfield  {journal} {\bibinfo  {journal} {Physics reports}\ }\textbf {\bibinfo {volume} {380}},\ \bibinfo {pages} {235} (\bibinfo {year} {2003})}\BibitemShut {NoStop}%
\bibitem [{\citenamefont {Zare}(1991)}]{zare}%
  \BibitemOpen
  \bibfield  {author} {\bibinfo {author} {\bibfnamefont {R.~N.}\ \bibnamefont {Zare}},\ }\href@noop {} {\emph {\bibinfo {title} {Angular Momentum: Understanding Spatial Aspects in Chemistry and Physics}}},\ \bibinfo {edition} {1st}\ ed.\ (\bibinfo  {publisher} {Wiley},\ \bibinfo {year} {1991})\BibitemShut {NoStop}%
\bibitem [{\citenamefont {Berger}(1985)}]{berger}%
  \BibitemOpen
  \bibfield  {author} {\bibinfo {author} {\bibfnamefont {J.~O.}\ \bibnamefont {Berger}},\ }\href@noop {} {\emph {\bibinfo {title} {Statistical Decision Theory and Bayesian Analysis}}}\ (\bibinfo  {publisher} {Springer},\ \bibinfo {year} {1985})\BibitemShut {NoStop}%
\bibitem [{\citenamefont {{G{\'o}rski}}\ \emph {et~al.}(2005)\citenamefont {{G{\'o}rski}}, \citenamefont {{Hivon}}, \citenamefont {{Banday}}, \citenamefont {{Wandelt}}, \citenamefont {{Hansen}}, \citenamefont {{Reinecke}},\ and\ \citenamefont {{Bartelmann}}}]{healpix}%
  \BibitemOpen
  \bibfield  {author} {\bibinfo {author} {\bibfnamefont {K.~M.}\ \bibnamefont {{G{\'o}rski}}}, \bibinfo {author} {\bibfnamefont {E.}~\bibnamefont {{Hivon}}}, \bibinfo {author} {\bibfnamefont {A.~J.}\ \bibnamefont {{Banday}}}, \bibinfo {author} {\bibfnamefont {B.~D.}\ \bibnamefont {{Wandelt}}}, \bibinfo {author} {\bibfnamefont {F.~K.}\ \bibnamefont {{Hansen}}}, \bibinfo {author} {\bibfnamefont {M.}~\bibnamefont {{Reinecke}}},\ and\ \bibinfo {author} {\bibfnamefont {M.}~\bibnamefont {{Bartelmann}}},\ }\bibfield  {title} {\bibinfo {title} {{HEALPix: A Framework for High-Resolution Discretization and Fast Analysis of Data Distributed on the Sphere}},\ }\href {https://doi.org/10.1086/427976} {\bibfield  {journal} {\bibinfo  {journal} {\apj}\ }\textbf {\bibinfo {volume} {622}},\ \bibinfo {pages} {759} (\bibinfo {year} {2005})},\ \Eprint {https://arxiv.org/abs/astro-ph/0409513} {arXiv:astro-ph/0409513 [astro-ph]} \BibitemShut {NoStop}%
\bibitem [{\citenamefont {{Liu}}\ and\ \citenamefont {{Bunn}}(2016)}]{liubunn}%
  \BibitemOpen
  \bibfield  {author} {\bibinfo {author} {\bibfnamefont {H.}~\bibnamefont {{Liu}}}\ and\ \bibinfo {author} {\bibfnamefont {E.~F.}\ \bibnamefont {{Bunn}}},\ }\bibfield  {title} {\bibinfo {title} {{Fisher matrix optimization of cosmic microwave background interferometers}},\ }\href {https://doi.org/10.1103/PhysRevD.93.023512} {\bibfield  {journal} {\bibinfo  {journal} {\prd}\ }\textbf {\bibinfo {volume} {93}},\ \bibinfo {eid} {023512} (\bibinfo {year} {2016})},\ \Eprint {https://arxiv.org/abs/1511.03635} {arXiv:1511.03635 [astro-ph.CO]} \BibitemShut {NoStop}%
\bibitem [{\citenamefont {Ade}\ \emph {et~al.}(2016)\citenamefont {Ade}, \citenamefont {Aghanim}, \citenamefont {Arnaud}, \citenamefont {Arroja}, \citenamefont {Ashdown}, \citenamefont {Aumont}, \citenamefont {Baccigalupi}, \citenamefont {Ballardini}, \citenamefont {Banday}, \citenamefont {Barreiro} \emph {et~al.}}]{ade2016planck_inflation}%
  \BibitemOpen
  \bibfield  {author} {\bibinfo {author} {\bibfnamefont {P.}~\bibnamefont {Ade}}, \bibinfo {author} {\bibfnamefont {N.}~\bibnamefont {Aghanim}}, \bibinfo {author} {\bibfnamefont {M.}~\bibnamefont {Arnaud}}, \bibinfo {author} {\bibfnamefont {F.}~\bibnamefont {Arroja}}, \bibinfo {author} {\bibfnamefont {M.}~\bibnamefont {Ashdown}}, \bibinfo {author} {\bibfnamefont {J.}~\bibnamefont {Aumont}}, \bibinfo {author} {\bibfnamefont {C.}~\bibnamefont {Baccigalupi}}, \bibinfo {author} {\bibfnamefont {M.}~\bibnamefont {Ballardini}}, \bibinfo {author} {\bibfnamefont {A.}~\bibnamefont {Banday}}, \bibinfo {author} {\bibfnamefont {R.}~\bibnamefont {Barreiro}}, \emph {et~al.},\ }\bibfield  {title} {\bibinfo {title} {Planck 2015 results-xx. constraints on inflation},\ }\href@noop {} {\bibfield  {journal} {\bibinfo  {journal} {Astronomy \& Astrophysics}\ }\textbf {\bibinfo {volume} {594}},\ \bibinfo {pages} {A20} (\bibinfo {year} {2016})}\BibitemShut {NoStop}%
\bibitem [{\citenamefont {Vitenti}\ \emph {et~al.}(2019)\citenamefont {Vitenti}, \citenamefont {Peter},\ and\ \citenamefont {Valentini}}]{Vitenti_2019_patrick}%
  \BibitemOpen
  \bibfield  {author} {\bibinfo {author} {\bibfnamefont {S.~D.}\ \bibnamefont {Vitenti}}, \bibinfo {author} {\bibfnamefont {P.}~\bibnamefont {Peter}},\ and\ \bibinfo {author} {\bibfnamefont {A.}~\bibnamefont {Valentini}},\ }\bibfield  {title} {\bibinfo {title} {Modeling the large-scale power deficit with smooth and discontinuous primordial spectra},\ }\bibfield  {journal} {\bibinfo  {journal} {Physical Review D}\ }\textbf {\bibinfo {volume} {100}},\ \href {https://doi.org/10.1103/physrevd.100.043506} {10.1103/physrevd.100.043506} (\bibinfo {year} {2019})\BibitemShut {NoStop}%
\bibitem [{\citenamefont {Abazajian}\ \emph {et~al.}(2019)\citenamefont {Abazajian}, \citenamefont {Addison}, \citenamefont {Adshead}, \citenamefont {Ahmed}, \citenamefont {Allen}, \citenamefont {Alonso}, \citenamefont {Alvarez}, \citenamefont {Anderson}, \citenamefont {Arnold}, \citenamefont {Baccigalupi} \emph {et~al.}}]{abazajian2019cmbs4_science_book}%
  \BibitemOpen
  \bibfield  {author} {\bibinfo {author} {\bibfnamefont {K.}~\bibnamefont {Abazajian}}, \bibinfo {author} {\bibfnamefont {G.}~\bibnamefont {Addison}}, \bibinfo {author} {\bibfnamefont {P.}~\bibnamefont {Adshead}}, \bibinfo {author} {\bibfnamefont {Z.}~\bibnamefont {Ahmed}}, \bibinfo {author} {\bibfnamefont {S.~W.}\ \bibnamefont {Allen}}, \bibinfo {author} {\bibfnamefont {D.}~\bibnamefont {Alonso}}, \bibinfo {author} {\bibfnamefont {M.}~\bibnamefont {Alvarez}}, \bibinfo {author} {\bibfnamefont {A.}~\bibnamefont {Anderson}}, \bibinfo {author} {\bibfnamefont {K.~S.}\ \bibnamefont {Arnold}}, \bibinfo {author} {\bibfnamefont {C.}~\bibnamefont {Baccigalupi}}, \emph {et~al.},\ }\bibfield  {title} {\bibinfo {title} {Cmb-s4 science case, reference design, and project plan},\ }\href@noop {} {\bibfield  {journal} {\bibinfo  {journal} {arXiv preprint arXiv:1907.04473}\ } (\bibinfo {year} {2019})}\BibitemShut {NoStop}%
\bibitem [{\citenamefont {Shimon}\ \emph {et~al.}(2006)\citenamefont {Shimon}, \citenamefont {Rephaeli}, \citenamefont {O'Shea},\ and\ \citenamefont {Norman}}]{shimon2006cosmic_raphaeli}%
  \BibitemOpen
  \bibfield  {author} {\bibinfo {author} {\bibfnamefont {M.}~\bibnamefont {Shimon}}, \bibinfo {author} {\bibfnamefont {Y.}~\bibnamefont {Rephaeli}}, \bibinfo {author} {\bibfnamefont {B.}~\bibnamefont {O'Shea}},\ and\ \bibinfo {author} {\bibfnamefont {M.}~\bibnamefont {Norman}},\ }\bibfield  {title} {\bibinfo {title} {Cosmic microwave background polarization due to scattering in clusters},\ }\href@noop {} {\bibfield  {journal} {\bibinfo  {journal} {Monthly Notices of the Royal Astronomical Society}\ }\textbf {\bibinfo {volume} {368}},\ \bibinfo {pages} {511} (\bibinfo {year} {2006})}\BibitemShut {NoStop}%
\bibitem [{\citenamefont {Amblard}\ and\ \citenamefont {White}(2005)}]{amblard2005sunyaev_white}%
  \BibitemOpen
  \bibfield  {author} {\bibinfo {author} {\bibfnamefont {A.}~\bibnamefont {Amblard}}\ and\ \bibinfo {author} {\bibfnamefont {M.}~\bibnamefont {White}},\ }\bibfield  {title} {\bibinfo {title} {Sunyaev--zel’dovich polarization simulation},\ }\href@noop {} {\bibfield  {journal} {\bibinfo  {journal} {New Astronomy}\ }\textbf {\bibinfo {volume} {10}},\ \bibinfo {pages} {417} (\bibinfo {year} {2005})}\BibitemShut {NoStop}%
\bibitem [{\citenamefont {Mirmelstein}\ \emph {et~al.}(2020)\citenamefont {Mirmelstein}, \citenamefont {Shimon},\ and\ \citenamefont {Rephaeli}}]{mirmelstein2020detection}%
  \BibitemOpen
  \bibfield  {author} {\bibinfo {author} {\bibfnamefont {M.}~\bibnamefont {Mirmelstein}}, \bibinfo {author} {\bibfnamefont {M.}~\bibnamefont {Shimon}},\ and\ \bibinfo {author} {\bibfnamefont {Y.}~\bibnamefont {Rephaeli}},\ }\bibfield  {title} {\bibinfo {title} {Detection likelihood of cluster-induced cmb polarization},\ }\href@noop {} {\bibfield  {journal} {\bibinfo  {journal} {Astronomy \& Astrophysics}\ }\textbf {\bibinfo {volume} {644}},\ \bibinfo {pages} {A36} (\bibinfo {year} {2020})}\BibitemShut {NoStop}%
\bibitem [{\citenamefont {Shimon}\ \emph {et~al.}(2009)\citenamefont {Shimon}, \citenamefont {Rephaeli}, \citenamefont {Sadeh},\ and\ \citenamefont {Keating}}]{shimon2009power}%
  \BibitemOpen
  \bibfield  {author} {\bibinfo {author} {\bibfnamefont {M.}~\bibnamefont {Shimon}}, \bibinfo {author} {\bibfnamefont {Y.}~\bibnamefont {Rephaeli}}, \bibinfo {author} {\bibfnamefont {S.}~\bibnamefont {Sadeh}},\ and\ \bibinfo {author} {\bibfnamefont {B.}~\bibnamefont {Keating}},\ }\bibfield  {title} {\bibinfo {title} {Power spectra of cmb polarization by scattering in clusters},\ }\href@noop {} {\bibfield  {journal} {\bibinfo  {journal} {Monthly Notices of the Royal Astronomical Society}\ }\textbf {\bibinfo {volume} {399}},\ \bibinfo {pages} {2088} (\bibinfo {year} {2009})}\BibitemShut {NoStop}%
\bibitem [{\citenamefont {Khabibullin}\ \emph {et~al.}(2018)\citenamefont {Khabibullin}, \citenamefont {Komarov}, \citenamefont {Churazov},\ and\ \citenamefont {Schekochihin}}]{khabibullin2018polarization}%
  \BibitemOpen
  \bibfield  {author} {\bibinfo {author} {\bibfnamefont {I.}~\bibnamefont {Khabibullin}}, \bibinfo {author} {\bibfnamefont {S.}~\bibnamefont {Komarov}}, \bibinfo {author} {\bibfnamefont {E.}~\bibnamefont {Churazov}},\ and\ \bibinfo {author} {\bibfnamefont {A.}~\bibnamefont {Schekochihin}},\ }\bibfield  {title} {\bibinfo {title} {Polarization of sunyaev--zel'dovich signal due to electron pressure anisotropy in galaxy clusters},\ }\href@noop {} {\bibfield  {journal} {\bibinfo  {journal} {Monthly Notices of the Royal Astronomical Society}\ }\textbf {\bibinfo {volume} {474}},\ \bibinfo {pages} {2389} (\bibinfo {year} {2018})}\BibitemShut {NoStop}%
\bibitem [{\citenamefont {Schiappucci}\ \emph {et~al.}(2023)\citenamefont {Schiappucci}, \citenamefont {Bianchini}, \citenamefont {Aguena}, \citenamefont {Archipley}, \citenamefont {Balkenhol}, \citenamefont {Bleem}, \citenamefont {Chaubal}, \citenamefont {Crawford}, \citenamefont {Grandis}, \citenamefont {Omori} \emph {et~al.}}]{schiappucci2023measurement_of_tau}%
  \BibitemOpen
  \bibfield  {author} {\bibinfo {author} {\bibfnamefont {E.}~\bibnamefont {Schiappucci}}, \bibinfo {author} {\bibfnamefont {F.}~\bibnamefont {Bianchini}}, \bibinfo {author} {\bibfnamefont {M.}~\bibnamefont {Aguena}}, \bibinfo {author} {\bibfnamefont {M.}~\bibnamefont {Archipley}}, \bibinfo {author} {\bibfnamefont {L.}~\bibnamefont {Balkenhol}}, \bibinfo {author} {\bibfnamefont {L.}~\bibnamefont {Bleem}}, \bibinfo {author} {\bibfnamefont {P.}~\bibnamefont {Chaubal}}, \bibinfo {author} {\bibfnamefont {T.}~\bibnamefont {Crawford}}, \bibinfo {author} {\bibfnamefont {S.}~\bibnamefont {Grandis}}, \bibinfo {author} {\bibfnamefont {Y.}~\bibnamefont {Omori}}, \emph {et~al.},\ }\bibfield  {title} {\bibinfo {title} {Measurement of the mean central optical depth of galaxy clusters via the pairwise kinematic sunyaev-zel’dovich effect with spt-3g and des},\ }\href@noop {} {\bibfield  {journal} {\bibinfo  {journal} {Physical Review D}\ }\textbf {\bibinfo {volume} {107}},\ \bibinfo {pages} {042004} (\bibinfo {year}
  {2023})}\BibitemShut {NoStop}%
\bibitem [{\citenamefont {Flauger}\ \emph {et~al.}(2017)\citenamefont {Flauger}, \citenamefont {McAllister}, \citenamefont {Silverstein},\ and\ \citenamefont {Westphal}}]{flauger2017drifting_constraints_monodromy}%
  \BibitemOpen
  \bibfield  {author} {\bibinfo {author} {\bibfnamefont {R.}~\bibnamefont {Flauger}}, \bibinfo {author} {\bibfnamefont {L.}~\bibnamefont {McAllister}}, \bibinfo {author} {\bibfnamefont {E.}~\bibnamefont {Silverstein}},\ and\ \bibinfo {author} {\bibfnamefont {A.}~\bibnamefont {Westphal}},\ }\bibfield  {title} {\bibinfo {title} {Drifting oscillations in axion monodromy},\ }\href@noop {} {\bibfield  {journal} {\bibinfo  {journal} {Journal of Cosmology and Astroparticle Physics}\ }\textbf {\bibinfo {volume} {2017}}\bibinfo  {number} { (10)},\ \bibinfo {pages} {055}}\BibitemShut {NoStop}%
\end{thebibliography}%
\end{document}